\documentclass[aps,prb,twocolumn,superscriptaddress,showpacs,english]{revtex4-1}
\usepackage[T1]{fontenc}
\usepackage[latin9]{inputenc}
\usepackage{babel}
\usepackage{amsmath}
\usepackage{amssymb}
\usepackage{graphicx}
\usepackage{xcolor}
\makeatletter

\usepackage{babel}
\usepackage{upgreek}

\makeatother

\begin{document}
\global\long\def\LP#1#2{\varLambda_{#1}^{{\scriptscriptstyle \text{#2}}}}
 \global\long\def\FigInEq#1#2{\includegraphics[width=#1\textwidth]{#2}}

\title{Superconducting pairing mediated by spin-fluctuations from first
principles}

\author{F. Essenberger} 
\affiliation{Max Planck Institute of Microstructure Physics, Weinberg 2, D-06120 Halle, Germany.}
\author{A. Sanna}
\affiliation{Max Planck Institute of Microstructure Physics, Weinberg 2, D-06120 Halle, Germany.}
\author{A. Linscheid}
\affiliation{Max Planck Institute of Microstructure Physics, Weinberg 2, D-06120 Halle, Germany.}
\author{F. Tandetzky}
\affiliation{Max Planck Institute of Microstructure Physics, Weinberg 2, D-06120 Halle, Germany.}
\author{G. Profeta}
\affiliation{Consiglio Nazionale delle Ricerche - Superconducting and Innovative Materials and Devices (CNR-SPIN), 67100 L'Aquila, Italy}
\author{P. Cudazzo}
\affiliation{Nano-Bio Spectroscopy group,
  Dpto. F\'isica de Materiales, Universidad del Pa\'is Vasco,
  Centro de F\'isica de Materiales CSIC-UPV/EHU-MPC and DIPC,
  Av. Tolosa 72, E-20018 San Sebasti\'an, Spain}
\affiliation{European Theoretical Spectroscopy Facility (ETSF)}
\author{E.K.U. Gross}
\affiliation{Max Planck Institute of Microstructure Physics, Weinberg 2, D-06120 Halle, Germany.}

\begin{abstract}
We present the derivation of an \textit{ab-initio} and parameter free
effective electron-electron interaction that goes beyond the screened
RPA and accounts for superconducting pairing driven by spin-fluctuations.
The construction is based on many body perturbation theory and relies
on the approximation of the exchange-correlation part of the electronic
self-energy within time dependent density functional theory. This
effective interaction is included in an exchange correlation kernel
for superconducting density functional theory, in order to achieve
a completely parameter free superconducting gap equation. First results from applying the new functional to a simplified two-band electron gas model are consistent with experiments.

\end{abstract}

\date{\today }

\maketitle

\section{Introduction}

In the last 30 years the field of superconductivity has been revolutionized
by the discovery of high-temperature (hi-T$_{c}$) superconductivity
(SC). First the Cuprates were found in the 80s\citep{ZeitschriftPhysB.64.189,RevModPhys.60.585}
and then iron based compounds in the 2000s\citep{JACS.128.10012,Rep.Prog.Phys.74.124508,RevModPhys.83.1589}.
Numerous empirical/semi-empirical theoretical models have been developed
in order to grasp the essential physics of these materials\citep{Manske,PhysRevB.82.144510,RevModPhys.78.17}
and still the community is far from a general consensus on the origin
of the pairing mechanism. In our opinion consensus can only be achieved
in one single way: By developing a universal predictive theory of
(hi-T$_{c}$) SC that is fully parameter free and is able to reproduce
the essential properties of the SC (including its critical temperature,
complex gap function and excitation spectrum), under the only knowledge
of the atomic constituents and chemical structure.

Within the class of conventional (meaning phonon driven) SC, density
functional theory for the SC state (SCDFT)~\citep{OGK_SCDFT_PRL1988},
within the available functional\citep{PhysRevB.72.024545,SannaGross}, proved
to be predictive and reliable\cite{FlorisMgB2_PRL2005,Profeta_LiKAl_PRL2006,SannaCaC6_PRB2007,Floris_Pb_PRB2007,Cudazzo_H_PRL,Cudazzo_H_PRB_2010,MarquesSCDFTII2005,AkashiAritaPlasmon,Bersier_CaBeSi_PRB2010,Sanna_K_PRB2006}. 
However, since the pairing in the pnictides 
and cuprates is non-phononic\citep{PhysRevLett.101.026403,PhysRev.125.1263},
this SCDFT approach is not directly applicable, due to the limitations
of the functional.

In this work we carefully reconsider this functional and it's construction,
in particular the treatment of the electronic component of the pairing.
We aim to reach two goals in this work: The \textit{first} is very
general and not bound to SCDFT applications. We want to formulate
a screened effective electron-electron interaction that goes beyond
the $GW$ form and includes additional physical effects not present
in the random phase approximation (RPA) type of screening. 
In particular we aim to include the effect of low energy spin-fluctuations, in a
computationally feasible way and completely \textit{ab-initio} (\emph{i.e.}
without the use of parameters - like a Stoner exchange splitting). 
We focus on the spin-fluctuations because they are one of the prime 
candidates responsible for SC pairing in iron SC \citep{PhysRevLett.101.057003}; 
The \textit{second} goal we aim for is to cast this effective interaction
along the standard Coulomb and phonon contribution\citep{PhysRevB.72.024545}
in a functional that can be used within the \textit{ab-initio} SCDFT
framework.

The paper has the following outline: In the next section (Sec.\ \ref{sec:The-Status-Quo})
we discuss briefly the existing functionals of SCDFT. Then we propose
the set of relevant diagrams for representing the spin-fluctuations
(Sec.\ \ref{sub:Relevant-Diagrams}) and the corresponding self-energy
contribution are constructed in the Nambu formalism (Sec.\ \ref{sub:InvestigationParticleHole}).
After some additional approximation (Sec.\ \ref{sub:Local-Approximation})
the final form of the self-energy taking the spin-fluctuations into
account is presented in Sec.\ \ref{sec:FinalForm}. This self-energy
may be used also in many-body theory but the focus of this work lies
on SCDFT and hence in Sec.\ \ref{sec:Functional} a functional is
derived using the Sham-Schlüter connection. In the last part of the
present work (Sec.\ \ref{sec:TwoBandModel}) the functional is applied
to a two band model system and the trends with respect to the Coulomb,
phonon and spin-fluctuations (SF) contributions are investigated.

\section{A brief review of SCDFT \label{sec:The-Status-Quo}}

Before engaging in the task of constructing the effective interaction
and the corresponding functional, we briefly review the SCDFT framework
and the available functionals. SCDFT is based on a theorem of Oliveira,
Gross and Kohn\citep{OGK_SCDFT_PRL1988}, that extends the Hohenberg-Kohn
proof\citep{PhysRev.136.B864} of the 1-1 correspondence between density
and external potential to the SC density 
\[
\chi\left(\boldsymbol{r},\!\boldsymbol{r}'\right)=\langle\hat{\varPsi}_{\downarrow}(\boldsymbol{r})\hat{\varPsi}_{\uparrow}(\boldsymbol{r}')\rangle
\]
where $\hat{\Psi}_{\sigma}\left(\boldsymbol{r}\right)$ are the usual
electronic field operators and $\left\langle \right\rangle $ is the
thermal average. The modern version of the theory has been re-formulated
by Lüders, Marques and coworkers\citep{PhysRevB.72.024545,MarquesSCDFTII2005}.
This formulation includes an explicit ionic density and a further
extension of the Hohenberg-Kohn proof in the spirit of the multicomponent
DFT introduced by Kreibich and Gross\citep{PhysRevA.78.022501}.

In their work, Lüders, Marques and coworkers\citep{PhysRevB.72.024545,MarquesSCDFTII2005}
proposed an exchange-correlation functional derived from many-body
perturbation theory and presented solutions of the SC Kohn-Sham (KS)
system for real SC. The starting point is an approximation for the
self-energy. In their work they use: 
\begin{align}
\bar{\varSigma}_{k}\left(\omega_{n}\right) & \approx\sum_{m}\sum_{k'}W_{kk'}\left(\omega_{n}-\omega_{m}\right)\bar{G}_{k'}^{{\scriptscriptstyle \text{KS}}}\left(\omega_{m}\right)\label{eq:SigmaEl}\\
 & +\sum_{m}\sum_{k'}\LP{kk'}{Ph}\left(\omega_{n}-\omega_{m}\right)\bar{G}_{k'}^{{\scriptscriptstyle \text{KS}}\text{ }}\left(\omega_{m}\right),\label{eq:SigmaPh}
\end{align}
where $\bar{G}_{k}^{\text{\ensuremath{{\scriptscriptstyle KS}}}}\left(\omega_{n}\right)$
is the Green's function of the SC Kohn-Sham system, in Nambu notation
\citep{PhysRev.117.648}, $W_{kk'}\left(\omega_{n}\right)$ is the
screened Coulomb interaction and $\LP{kk'}{Ph}$ is the interaction
mediated by phonons. We will indicate objects in Nambu notation with
a bar (for example $\bar{G}$). The components of the Green's
function read: 
\begin{equation}
\bar{G}_{k}\left(\omega_{n}\right)=\uptau^{\mathrm{z}}\begin{pmatrix}G_{k}\left(\omega_{n}\right) & F_{k}\left(\omega_{n}\right)\\
{F_{k}}^{\dagger}\left(\omega_{n}\right) & {G_{k}}^{\dagger}\left(\omega_{n}\right)
\end{pmatrix}.
\end{equation}
The $\omega_{n}$ are the fermionic Matsubara frequencies, $k$ is
a combined index $k=\left\{ n\boldsymbol{k}\right\} $ containing
the band index and the momentum of the KS electron and $\uptau^{\mathrm{z}}$
is the third Pauli matrix. The normal ($G_{k}$) and anomalous ($F_{k})$
part of the Nambu Green's function are given by: 
\begin{align*}
G_{k}\left(\omega_{n}\right) & =-\int_{0}^{\beta}\mathrm{d}\tau\mathrm{e}^{\mathrm{i}\omega_{n}\tau}\left\langle \hat{\mathrm{T}}\left[\hat{a}_{k}\left(\tau\right)\hat{a}_{k}^{\dagger}\left(0\right)\right]\right\rangle \\
F_{k}\left(\omega_{n}\right) & =-\int_{0}^{\beta}\mathrm{d}\tau\mathrm{e}^{\mathrm{i}\omega_{n}\tau}\left\langle \hat{\mathrm{T}}\left[\hat{a}_{k}\left(\tau\right)\hat{a}_{k}\left(0\right)\right]\right\rangle 
\end{align*}
where $\hat{a}_{k}\left(\tau\right)$ and $\hat{a}_{k}^{\dagger}$
are the usual creation and annihilation operators in the Heisenberg
picture, $\hat{\mathrm{T}}$ is the time ordering operator and $\left\langle \right\rangle $
denotes the thermal average. The electronic part of the interaction,
in the work of Marques and Lüders, is assumed to be given by the classical
(test-charge to test-charge) screened Coulomb interaction\citep{PhysRevB.20.550},
therefore it can be expressed in terms of the dielectric function
$\epsilon^{-1}$: 
\begin{equation}
W_{k_{1}k_{2}}\left(\omega_{n}\right)=\sum_{k'}\epsilon_{k_{1}k'}^{-1}\left(\omega_{n}\right)v_{k'k_{2}}\label{eq:dielectric}
\end{equation}
where $v_{k_{1}k_{2}}$ is the bare Coulomb interaction. The 
interaction mediate by phonons $\LP{kk'}{Ph}\left(\omega_{n}\right)$ depends on the
electron-phonon coupling matrix elements $g_{\lambda\boldsymbol{q}}^{kk'}$
and the phonon frequencies $\varOmega_{\lambda\boldsymbol{q}}$: 
\begin{align}
\LP{kk'}{Ph}\left(\omega_{n}\right) & =-\frac{1}{\uppi}\int_{0}^{\infty}\!\!\!\mathrm{d}\omega\frac{2\omega}{\omega_{n}^{2}+\omega^{2}}\mathfrak{Im}\left[\LP{kk'}{Ph}\left(\omega\right)\right]\label{eq:KKrelation}\\
\mathfrak{Im}\left[\LP{kk'}{Ph}\left(\omega\right)\right] & =-\uppi\sum_{\lambda\boldsymbol{q}}\left|g_{\lambda\boldsymbol{q}}^{kk'}\right|^{2}\updelta\left(\omega-\varOmega_{\lambda\boldsymbol{q}}\right).\nonumber 
\end{align}
A Feynman diagram schematic form for this approximation is shown in
Eq.\ \ref{eq:statusquo}. For the two terms we use the names $\bar{\varSigma}^{{\scriptscriptstyle GW}}$
and $\bar{\varSigma}^{{\scriptscriptstyle \text{Ph}}}$, respectively.

\begin{minipage}[t]{1\columnwidth}%
\begin{center}

\par\end{center}

\begin{center}
%\begin{gather}
\FigInEq{1.0}{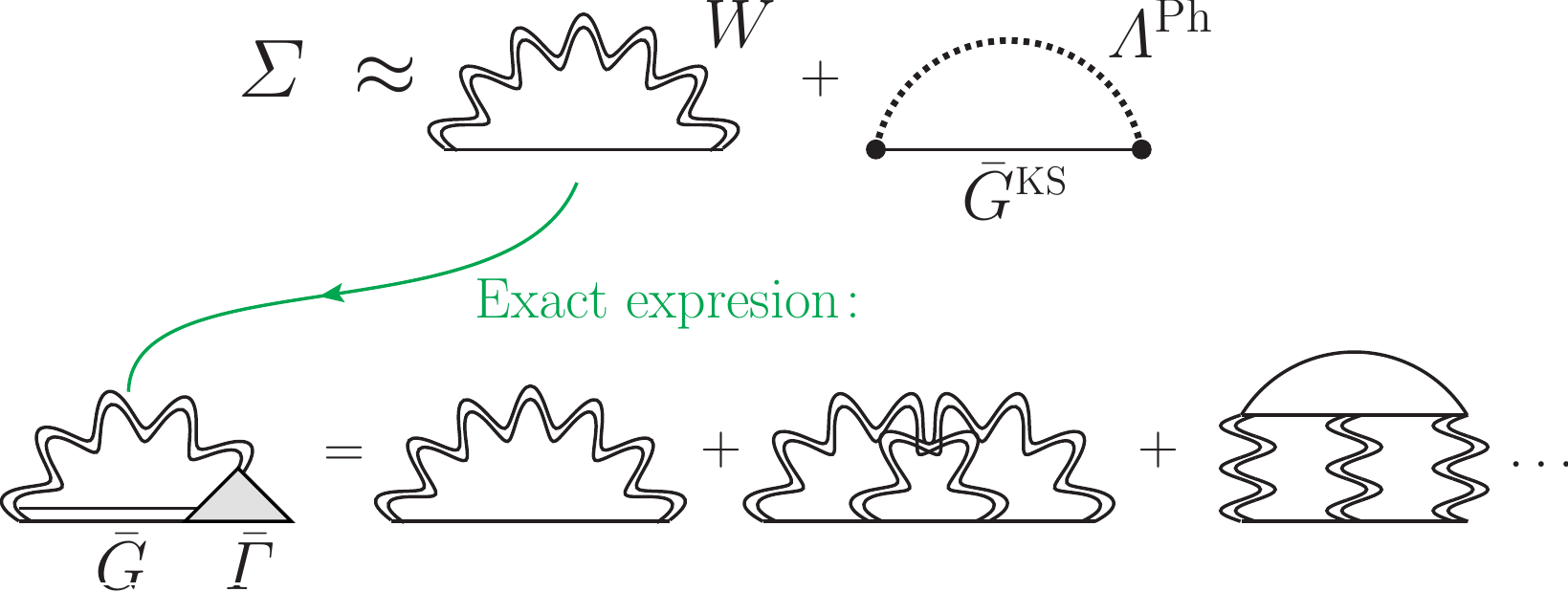}\raisetag{50pt}\label{eq:statusquo}
%\end{gather}

\par\end{center}%
\end{minipage}

Once an approximation for the self-energy is fixed, it is possible
to construct the corresponding exchange-correlation (xc) potential
using the Sham-Schlüter connection (see Sec.\ \ref{sec:Functional}
for details). A key approximation in order to reduce the numerical
complexity of SCDFT is the so-called decoupling approximation\cite{GrossKurth1991} \emph{i.e.}
\[
\varDelta_{kk'}^{\text{xc}}\approx\updelta_{kk'}\varDelta_{k}^{\text{xc}}
\]
that can be interpreted as the exclusion of hybridization effects
between the non SC Kohn-Sham orbitals by the effect of the SC condensation.
In this approximation the electronic KS system can be diagonalized
analytically leading to a self-consistent expression for the pairing
potential $\varDelta_{k}^{\text{xc}}$ known as the SCDFT gap equation:
\begin{equation}
\varDelta_{k}^{\text{xc}}=-\varDelta_{k}^{\text{xc}}\mathcal{Z}_{k}^{\mathrm{{\scriptscriptstyle D}}}-\sum_{k'}\mathcal{K}_{kk'}^{\mathrm{{\scriptscriptstyle C}}}\frac{\tanh\left(\frac{\beta E_{k'}}{2}\right)}{2E_{k'}}\varDelta_{k'}^{\text{xc}}\label{eq:gap}
\end{equation}
that has the BCS form\citep{PhysRev.108.1175}. The kernels $\mathcal{Z}_{k}^{\mathrm{{\scriptscriptstyle D}}}$
and $\mathcal{K}_{kk'}^{\mathrm{{\scriptscriptstyle C}}}$ depend
on the temperature, the interaction matrix elements $\left(w_{k_{1}k_{2}},\LP{kk'}{Ph}\right)$,
the single particle KS energies $\epsilon_{k}$ and $E_{k}:=\sqrt{\left|\varDelta_{k}\right|^{2}+\left(\epsilon_{k}-\mu\right)^{2}}$.
The critical temperatures predicted within this equation agree extremely
well with the experimentally observed ones within the class of phononic
SC\cite{MarquesSCDFTII2005,FlorisMgB2_PRL2005,Floris_Pb_PRB2007,SannaCaC6_PRB2007,Profeta_LiKAl_PRL2006,Gonnelli_CaC6_PRL2008,Sanna_K_PRB2006,Bersier_PbHCaBeSi_JOP2009}.
However, Eq.~\ref{eq:gap} fails to describe high temperature SC\citep{RevModPhys.83.1589,RevModPhys.72.969},
where the SC mechanism is believed to be related to magnetic interactions.

In the next sections we will see that this fact is actually not surprising.
One assumption in using a dielectric type of electron-electron interaction
is that all vertex corrections in the Coulomb part of the self-energy
are completely neglected. As one can see in Eq.\ \ref{eq:statusquo},
by comparing the approximation with its exact counterpart obtained
from Hedin's equations\citep{PhysRev.139.A796}. Vertex corrections
can be safely disregarded in the phonon related part of the self-energy
(at least within the domain of validity of Migdal's theorem\citep{JETP.34.996}),
but are crucial to account for magnetic fluctuation effects which
will be discussed in the next sections.

\section{Extension of the Self-Energy\label{sec:Extension_of_Sigma}}

In this section we will construct a form of the self-energy containing
the relevant processes involved in a spin-fluctuation-mediated pairing.
The effective interaction will be evaluated in the parent metallic system
in which SC takes place (\emph{i.e.} we will ignore the feedback effect
of the SC condensation). This approximation may not be valid at low
temperature where the condensation strongly affects the screening
of magnetic fluctuations\citep{PhysRevB.79.134520,PhysRevB.81.100513,Manske}.
However this assumption is exact near the critical point since the
SC phase transition is of the continuous, second order type. Therefore
the approximation will not affect the estimation of a critical temperature.

Note, that the same approximation was applied to the phononic part
of the interaction, entering the gap equation (Eq.~\ref{eq:gap}).
In this case the effect of the condensation on the pairing strength
is most likely negligible even at low temperature\citep{PhysRevB.12.4899,PhysRev.129.2495}.

\subsection{Inclusion of the Relevant Diagrams\label{sub:Relevant-Diagrams}}

To go beyond the $GW$ approximation we consider
\footnote{A more general and unbiased  way would be 
to start from Hedin cycle\citep{PhysRev.139.A796} and iterate it self-consistently. This would lead also include the $T$-matrix diagrams considered here, but at a slow convergence rate\cite{PhysRev.139.A796}.}
 the $T$-matrix\cite{PhysRevB.81.054434,RevModPhys.74.601}, that is given
by a Bethe-Salpeter equation (BSE) \citep{PhysRev.84.1232}: 
\begin{align}
T\left(\mathit{1,\!2,\!3,\!4}\right) & =w\left(\mathit{1,\!3}\right)\updelta_{\mathit{13}}\updelta_{\mathit{24}}\label{eq:TmatAnalytsich}\\
 & +w\left(\mathit{1,\!2}\right)G\left(\mathit{1,\!5}\right)G\left(\mathit{2,\!6}\right)T\left(\mathit{5,\!6,\!3,\!4}\right).\nonumber 
\end{align}
The coordinate $\mathit{1}$ is a compact notation: $\mathit{1}=\left\{ \boldsymbol{r}_{\mathit{1}},\!\tau_{\mathit{1}},\!\sigma_{\mathit{1}}\right\} $,
where $\boldsymbol{r}_{1}$ is the real space vector, $\tau_{1}$
the Matsubara time and $\sigma_{1}$ the spin index. The diagrammatic
form of this BSE and the self-energy contribution $\bar{\varSigma}^{{\scriptscriptstyle T}}=\bar{G}T$
corresponding to the $T-$matrix are shown in Eq.~\ref{eq:Tmat}.

\begin{minipage}[t]{0.9\columnwidth}%

\begin{center}
%\begin{gather}
\includegraphics[width=1.18\columnwidth]{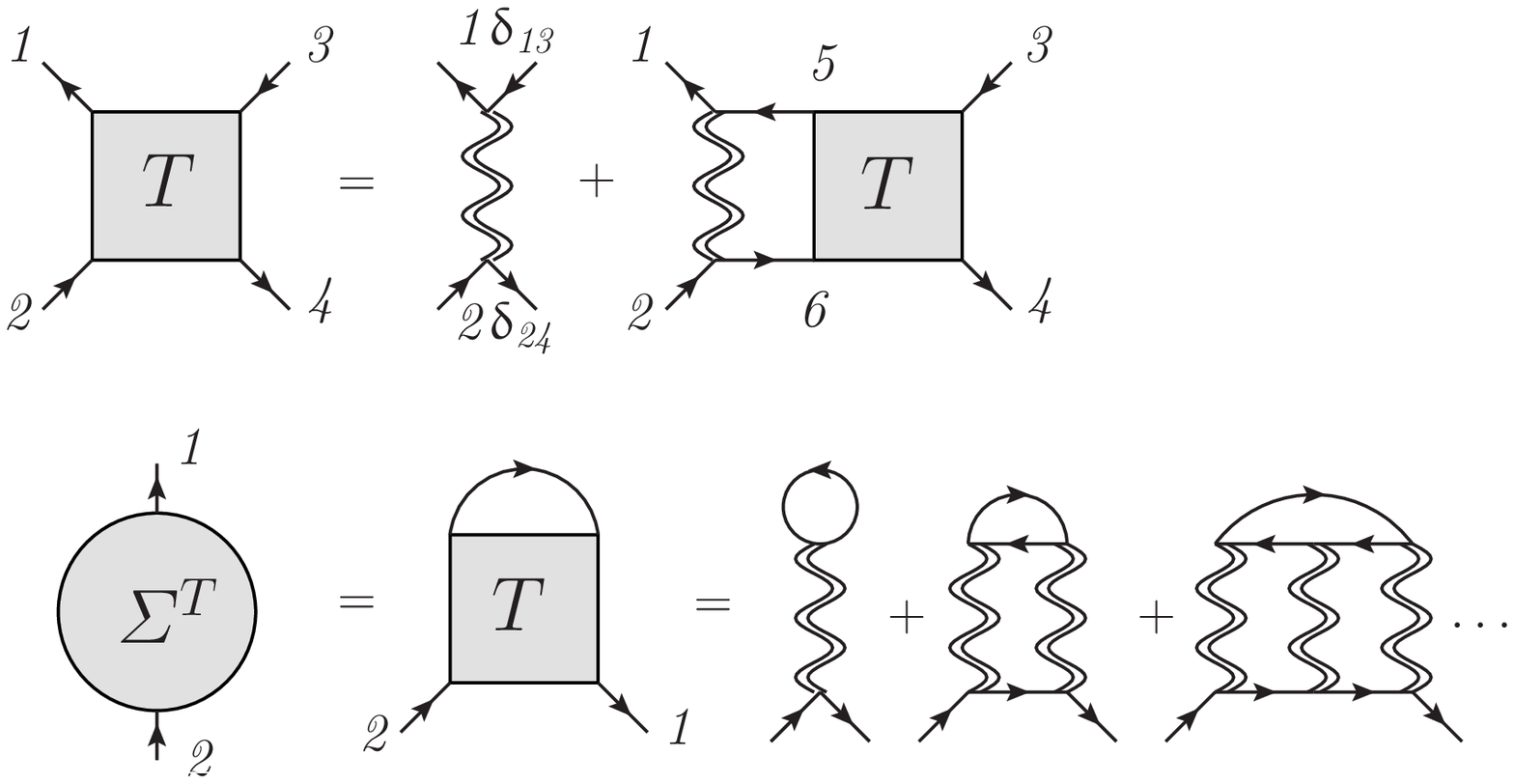}\raisetag{80pt}\label{eq:Tmat}
%\end{gather}

\par\end{center}%
\end{minipage}

Empirically it is well known, that the response function in the $T$-matrix
approximation leads to reasonable results for the magnetic response
function \citep{PhysRevB.81.054434,RevModPhys.74.601}. $\bar{\varSigma}^{{\scriptscriptstyle T}}$
has been used in various studies to account for magnetic fluctuations
in non-SC systems \citep{JoPF.3.2174,PhysRevLett.17.750,PhysRevB.72.155109,Romaniello_GW_PRB2012}.

However, for reasons that we will discuss in Sec.~\ref{sub:Local-Approximation},
we will not make direct use of the $T$-matrix and the corresponding
self-energy for constructing the effective interaction. Instead we
will consider a larger set of diagrams, by starting from the particle-hole
propagator $\varLambda^{{\scriptscriptstyle \text{P}}}$ \citep{PhysRevB.32.2156,PhysRevB.71.113102,PhysRevLett.100.116402}.
This object contains all proper particle-hole contributions. These
are all diagrams which are irreducible with respect to a bare Coulomb
interaction and have two incoming and two outgoing open coordinates.
The $T-$matrix is fully contained in $\varLambda^{{\scriptscriptstyle \text{P}}}$.
We use the analogy with $\bar{\varSigma}^{{\scriptscriptstyle T}}$ (see Eq.~\ref{eq:Tmat}),
to formulate the self-energy containing magnetic fluctuations as:

\begin{minipage}[t]{1\columnwidth}%

\begin{center}
%\begin{gather}
\FigInEq{1.0}{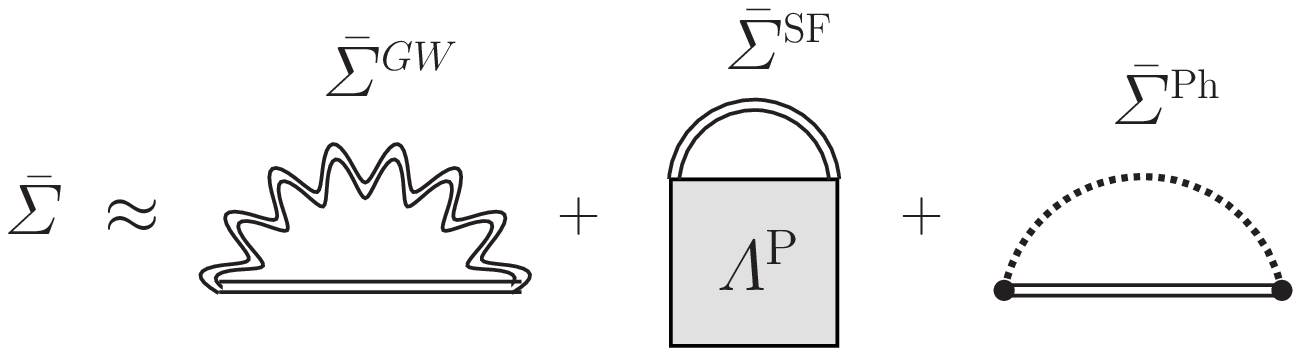}\raisetag{40pt}\label{eq:adhoc}
%\end{gather}

\par\end{center}%
\end{minipage}

In Eq.~\ref{eq:adhoc} we only show a simple diagrammatic form, details
will be derived explicitly in the next section. Note that this form
of the self-energy contains both Hartree and xc contributions 
 while only the xc parts enter the functional
derivative appearing in the vertex part of Hedin's equations. The
Hartree contribution will be implicitly removed in Sec.~\ref{sub:Local-Approximation}
when we define the approximation for $\varLambda^{{\scriptscriptstyle \text{P}}}$.
Also double counting problems related to this choice of the self-energy
are adressed in Sec.~\ref{sub:Local-Approximation}. As a general
convention in this work we will always refer to the $xc$ (Hartree
free) part of the self-energy.

\subsection{Properties of the Particle-Hole Propagator\label{sub:InvestigationParticleHole}}

In this section we will investigate the properties of the particle-hole
propagator, which is the key object of our derivation. For simplicity we will restrict ourselves to collinear magnetic systems, \emph{ i.e.} we assume a spin-diagonal Green function $G\left(\mathit{1,\!2}\right)=\updelta_{\sigma_{1}\sigma_{2}}G\left(\mathit{1,\!2}\right)$.
One of Hedin's equations is a Dyson equation for the vertex $\varGamma\left(\mathit{1,\!2,\!3}\right)$:

\begin{minipage}[t]{0.9\columnwidth}%

\begin{center}
%\begin{gather}
\FigInEq{1.0}{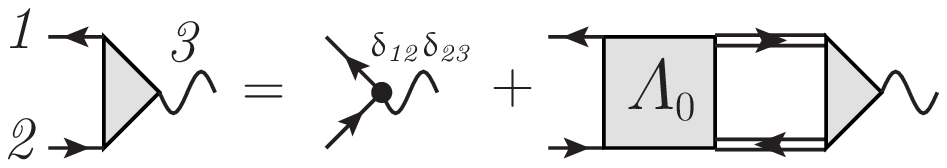}\raisetag{10pt}\label{eq:Vertex}
%\end{gather}

\par\end{center}

\end{minipage}

where the kernel of the Dyson equation is given by 
\begin{equation}
\varLambda_{0}\left(\mathit{1,\!2,\!3,\!4}\right):=\frac{\updelta\varSigma^{{\scriptscriptstyle \text{V}}}\left(\mathit{1,\!2}\right)}{\updelta G\left(\mathit{3,\!4}\right)}\label{eq:DefLambda0}
\end{equation}
 and is called irreducible particle-hole propagator \citep{PhysRevLett.100.116402}.
The $\varLambda_{0}$ contains all connected diagrams which are irreducible
with respect to a bare Coulomb interaction and the particle-hole propagator.
The coordinates $\mathit{1}$ and $\mathit{4}$ are connected to outgoing
Green's functions and $\mathit{2}$ and $\mathit{3}$ to incoming
ones. (Eq.~\ref{eq:DirectCrossed}). The self-energy used in the
construction of the kernel will be indicated by $\varSigma^{{\scriptscriptstyle \text{V}}}$.
The kernel $\varLambda_{0}$ also plays the central role in the BSE
equation for $\varLambda^{{\scriptscriptstyle \text{P}}}$ which
we shall derive now. However, before this can be done it is necessary
to classify the two possible contribution present in $\varLambda_{0}$.
The distinction between the two sets is made using the concept of a path.
A path is a chain of Green's function lines connecting to coordinates.
For example in Eq.\ \ref{eq:Path}, we have a path connecting the
coordinates 1 and 4:

\begin{minipage}[t]{0.9\columnwidth}%

\begin{center}
%\begin{gather}
\includegraphics[width=0.6\columnwidth]{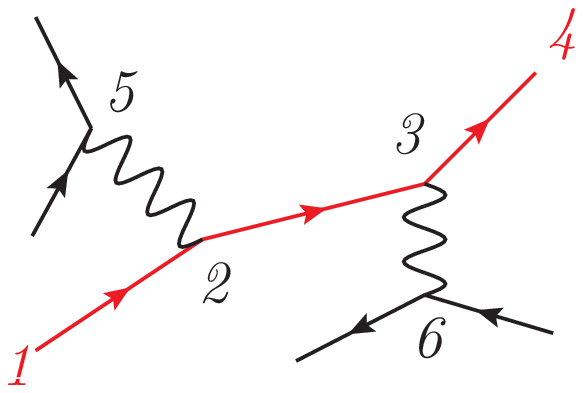}\raisetag{82pt}\label{eq:Path}
%\end{gather}

\par\end{center}%
\end{minipage}

After this definition we can introduce the two possible contribution
present in $\varLambda_{0}$:\vspace{0.2cm}

\textbf{1)} The crossed contribution $\varLambda_{0}^{\text{c}}$,
which has a path connecting the coordinates $\mathit{1}\leftrightarrow\mathit{3}$
and $\mathit{2}\leftrightarrow\mathit{4}$. The spin contributions
in this set are 
\begin{equation}
\varLambda_{0}^{\text{c}}\left(\mathit{1,\!2,\!3,\!4}\right)\equiv\updelta_{\sigma_{\mathit{1}}\sigma_{\mathit{3}}}\updelta_{\sigma_{\mathit{2}}\sigma_{\mathit{4}}}\varLambda_{0}^{\text{c}}\left(\mathit{1,\!2,\!3,\!4}\right).\label{eq:ChangeExact-SpinLambdaC}
\end{equation}
Note, that the contributions to the $T$-matrix (Eq.~\ref{eq:Tmat})
are all of this type \citep{PhysRevB.81.054434}.\vspace{0.2cm}

\textbf{2)} The direct contribution $\varLambda{}_{0}^{\text{d}}$,
which has a path connecting the coordinates $\mathit{1}\leftrightarrow\mathit{2}$
and $\mathit{3}\leftrightarrow\mathit{4}$. The spin contributions
in this set are 
\begin{equation}
\varLambda_{0}^{\text{d}}\left(\mathit{1,\!2,\!3,\!4}\right)\equiv\updelta_{\sigma_{\mathit{1}}\sigma_{\mathit{2}}}\updelta_{\sigma_{\mathit{3}}\sigma_{\mathit{4}}}\varLambda_{0}^{\text{d}}\left(\mathit{1,\!2,\!3,\!4}\right).\label{eq:ChangeExact-SpinLambdaD}
\end{equation}
The kernels $\varLambda_{0}^{\text{c}}$ and $\varLambda_{0}^{\text{d}}$
are created by the functional derivative of the self-energy with respect
to $G$. By the functional derivative $\frac{\updelta}{\updelta G\left(\mathit{3,\!4}\right)}$
one Green's function within the self-energy is removed and the open
connections get the indices 3 and 4 resulting in the four-point function
$\varLambda_{0}\left(\mathit{1,\!2,\!3,\!4}\right)$.

\begin{minipage}[t]{0.9\columnwidth}%

\begin{center}
%\begin{gather}
\includegraphics[width=1.18\columnwidth]{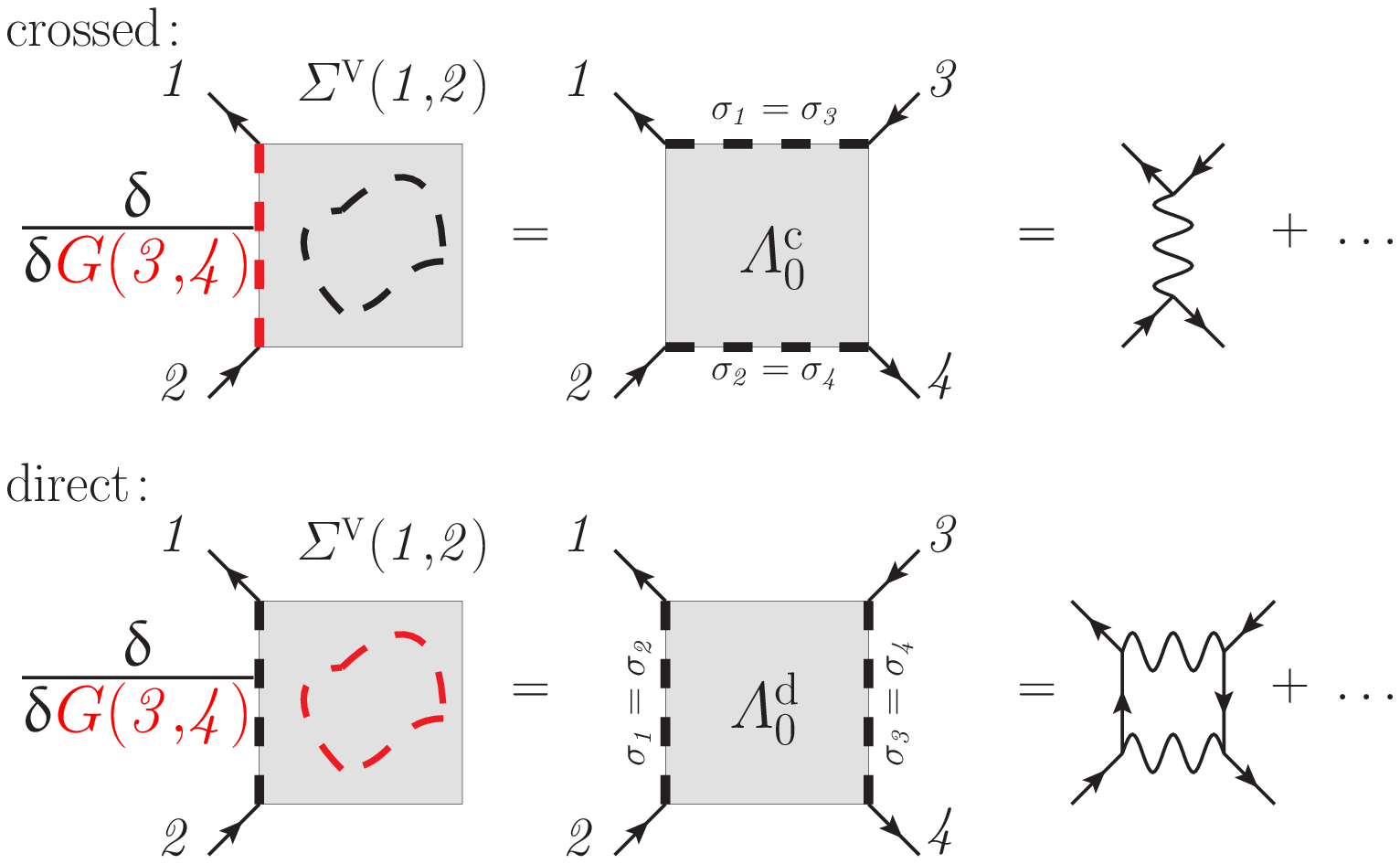}\raisetag{82pt}\label{eq:DirectCrossed}
%\end{gather}

\par\end{center}%
\end{minipage}

If the removed function was part of a loop, the resulting contribution
is direct. It is crossed otherwise (Eq.~\ref{eq:DirectCrossed}).
Since a loop was destroyed in the derivative, an extra minus sign
is necessary to compensate for this:

\begin{align}
\varLambda_{0}^{\text{c}}\left(\mathit{1,\!2,\!3,\!4}\right) & =\frac{\updelta\varSigma^{{\scriptscriptstyle \text{V}}}\left(\mathit{1,\!2}\right)}{\updelta G\left(\mathit{3,\!4}\right)}\text{ with }G\text{ not in loop}\label{eq:DerivativeNotLoop}\\
\varLambda_{0}^{\text{d}}\left(\mathit{1,\!2,\!3,\!4}\right) & =-\frac{\updelta\varSigma^{{\scriptscriptstyle \text{V}}}\left(\mathit{1,\!2}\right)}{\updelta G\left(\mathit{3,\!4}\right)}\text{ with }G\text{ in loop}.\label{eq:DerivativeInLoop}
\end{align}
It is important to keep track for these \textit{signs} since, while
Feynman diagrams have an explicit sign convention, symbolic expressions
(like the ones written in terms of the particle hole propagator Eq.\ \ref{eq:adhoc})
do not.

Due to Eqs.~\ref{eq:DefLambda0},\ref{eq:DerivativeNotLoop} and \ref{eq:DerivativeInLoop},
the total irreducible particle-hole propagator is given by the difference
between the crossed and direct contributions: 
\begin{equation}
\varLambda_{0}\left(\mathit{1,\!2,\!3,\!4}\right)=\frac{\updelta\varSigma^{{\scriptscriptstyle \text{V}}}\left(\mathit{1,\!2}\right)}{\updelta G\left(\mathit{3,\!4}\right)}=\varLambda_{0}^{\text{c}}-\varLambda_{0}^{\text{d}}=:\varLambda_{0}^{\text{c}{\scriptscriptstyle -}\mathrm{d}}.\label{eq:Lambda_0_all}
\end{equation}
With these preliminary considerations we can start to derive a BSE
for $\varLambda^{{\scriptscriptstyle \text{P}}}$. Note that also
within the set $\varLambda^{{\scriptscriptstyle \text{P}}}$ all contributions
are either direct or crossed, \emph{i.e.} $\varLambda^{{\scriptscriptstyle \text{P}}}=\varLambda^{\text{c}}+\varLambda^{\text{d}}$.
If for example two crossed contribution are linked, the resulting
one stays crossed.

\begin{minipage}[t]{1\columnwidth}%

\begin{center}
%\begin{gather}
\begin{centering}\FigInEq{0.7}{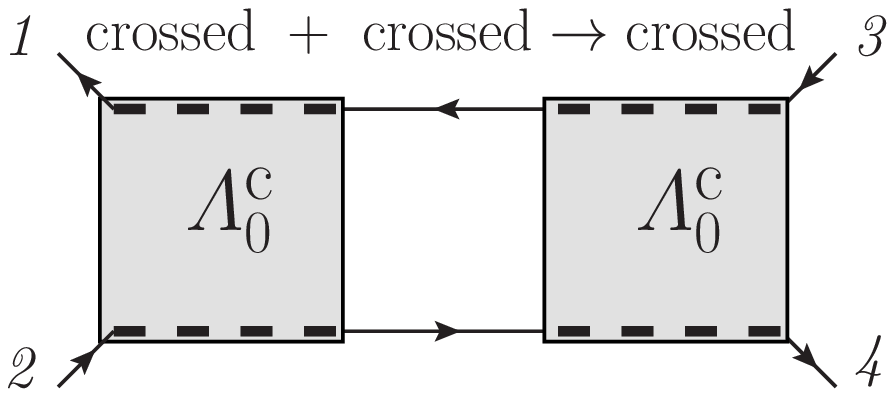}\end{centering}\raisetag{20pt}\label{eq:crossedChain}
%\end{gather}

\par\end{center}%
\end{minipage}

Any other combination leads to a direct contribution. A special case
is the connection of two direct contribution, in which a loop is created:

\begin{minipage}[t]{1\columnwidth}%

\begin{center}
%\begin{gather}
\begin{centering}\FigInEq{0.7}{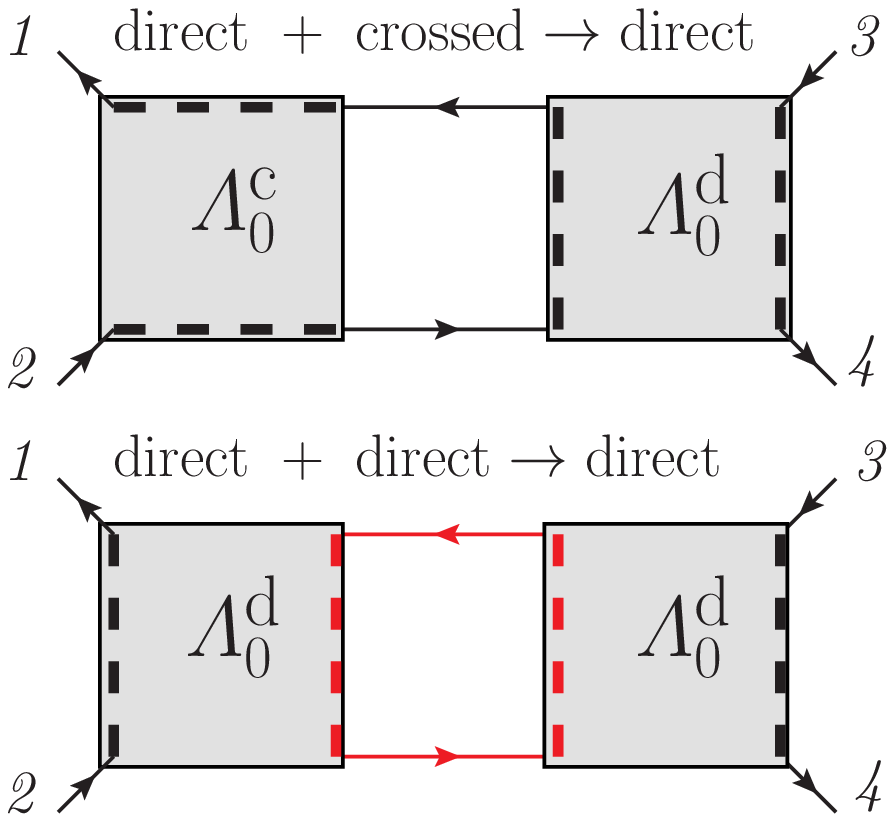}\end{centering}\raisetag{20pt}\label{eq:directChain}
%\end{gather}

\par\end{center}%
\end{minipage}

Considering these cases, the BSEs for the direct and crossed contribution
of the particle-hole propagator read: 
\begin{align}
\varLambda^{{\scriptscriptstyle \text{P}}} & =\sum_{n=0}^{\infty}\varLambda_{\left(n\right)}^{\text{\text{c}}}+\sum_{n=0}^{\infty}\varLambda_{\left(n\right)}^{\text{\text{d}}}\nonumber \\
\varLambda_{\left(n+1\right)}^{\text{c}} & =\varLambda_{0}^{\text{c}}GG\varLambda_{\left(n\right)}^{\text{c}}\label{eq:LambdaCsep}\\
\varLambda_{\left(n+1\right)}^{\text{d}} & =\varLambda_{0}^{\text{d}}GG\varLambda_{\left(n\right)}^{\text{c}}+\varLambda_{0}^{\text{c}}GG\varLambda_{\left(n\right)}^{\text{d}}-\varLambda_{0}^{\text{d}}GG\varLambda_{\left(n\right)}^{\text{d}}.\label{eq:LambdaDsep}
\end{align}
where $(n)$ labels the order in the irreducible particle-hole propagator
and the zero order $\varLambda_{\left(0\right)}^{\text{c,d}}$ is
given by the irreducible part $\varLambda_{0}^{\text{c,d}}$. By subtracting
 Eqs.\ \ref{eq:LambdaCsep} and \ref{eq:LambdaDsep} we find a
combined BSE for $\varLambda^{\text{c-d}}$ containing crossed and
direct terms: 
\begin{equation}
\varLambda^{\text{c}{\scriptscriptstyle -}\mathrm{d}}=\varLambda_{0}+\varLambda_{0}GG\varLambda^{\text{c}{\scriptscriptstyle -}\mathrm{d}}\text{ with }\varLambda_{0}=\frac{\updelta\varSigma^{{\scriptscriptstyle \text{V}}}}{\updelta G}.\label{eq:RelevantBSE}
\end{equation}
Not only in the BSE, also for the expression for $\bar{\varSigma}^{{\scriptscriptstyle \text{SF}}}$
given in Eq.\ \ref{eq:adhoc} the separation in direct and crossed
contribution is crucial. Up to now only the normal state Green's function
appeared in the equations, because we neglected the feedback effects
of SC to the magnetic fluctuations. However, in the expression for
the self-energy (Eq.\ \ref{eq:adhoc}) the normal and anomalous parts
appear and double arrow lines

\begin{minipage}[t]{1\columnwidth}%

\begin{center}
%\begin{gather}
\begin{centering}\FigInEq{0.9}{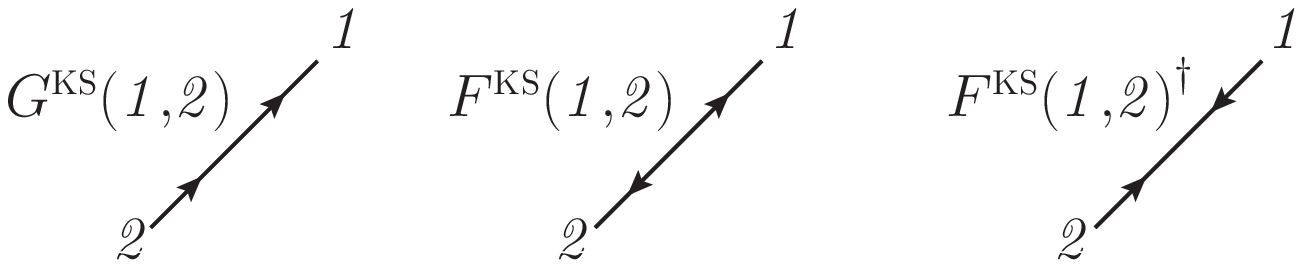}\end{centering}\raisetag{20pt}\label{eq:DefG_F}
%\end{gather}

\par\end{center}%
\end{minipage}

are used to distinguish the different functions. Since in the anomalous
terms no extra loops are created

\begin{minipage}[t]{1\columnwidth}%

\begin{center}
%\begin{gather}
\begin{centering}\FigInEq{0.8}{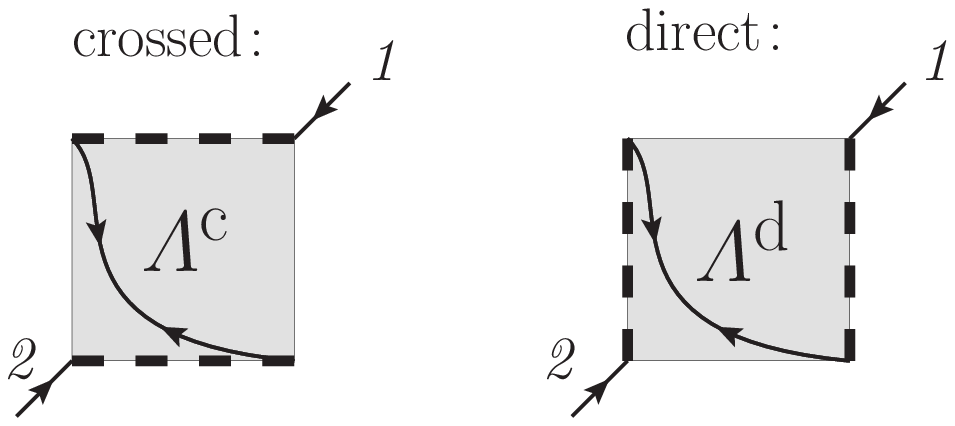}\end{centering}\raisetag{20pt}\label{eq:LoopRule-1}
%\end{gather}

\par\end{center}%
\end{minipage}the crossed and direct contributions enter both with the same sign
in the equation for the self-energy: 
\begin{equation}
\bar{\varSigma}_{\text{\ensuremath{{\scriptscriptstyle F}}}}^{{\scriptscriptstyle \text{SF}}}:=\iint\uptau^{\mathrm{z}}\begin{pmatrix}0 & F\varLambda^{\text{c}{\scriptscriptstyle +}\text{d}}\\
{F}^{\dagger}\varLambda^{\text{c}{\scriptscriptstyle +}\text{d}} & 0
\end{pmatrix}.\label{eq:ChangeExact-LoopAnorm}
\end{equation}
For the normal contribution (diagonal component in Nambu space) the
situation is a bit more complicated because the loop rule has to
be taken into account: If a crossed contribution is inserted inside the self-energy
form in Eq.~\ref{eq:adhoc}, then a loop is created (Fig.~\ref{eq:LoopRule})
leading to a minus sign. While the direct terms do not lead to any
additional loop and no sign change. This can be seen in the following
graph:

\begin{minipage}[t]{1\columnwidth}%

\begin{center}
%\begin{gather}
\begin{centering}\FigInEq{0.8}{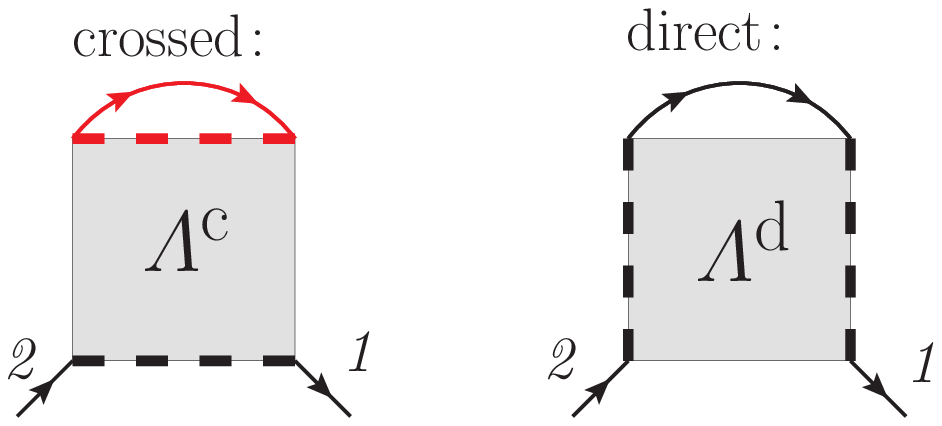}\end{centering}\raisetag{20pt}\label{eq:LoopRule}
%\end{gather}

\par\end{center}%
\end{minipage}

In the short hand notation given in Eq.\ \ref{eq:adhoc} this was
not, strictly speaking, taken into account. The rigorous form of this
equation instead reads:

\begin{align}
 & \bar{\varSigma}^{{\scriptscriptstyle \text{SF}}}:=\iint\mathrm{d}\mathit{34}\uptau^{\mathrm{z}}\begin{pmatrix}-G\left(\mathit{3,\!4}\right)\varLambda^{\text{c}{\scriptscriptstyle -}\mathrm{d}}\left(\mathit{1,\!3,\!2,\!4}\right) & 0\\
{F}^{\dagger}\left(\mathit{3,\!4}\right)\varLambda^{\text{c}{\scriptscriptstyle +}\text{d}}\left(\mathit{3,\!1,\!2,\!4}\right) & 0
\end{pmatrix}\label{eq:SigmaSF-1}\\
 & +\iint\mathrm{d}\mathit{34}\uptau^{\mathrm{z}}\begin{pmatrix}0 & F\left(\mathit{3,\!4}\right)\varLambda^{\text{c}{\scriptscriptstyle +}\text{d}}\left(\mathit{1,\!3,\!4,\!2}\right)\\
0 & -{G}^{\dagger}\left(\mathit{3,\!4}\right)\varLambda^{\text{c}{\scriptscriptstyle -}\mathrm{d}}\left(\mathit{3,\!1,\!4,\!2}\right)
\end{pmatrix}.\nonumber 
\end{align}
Here it appears explicitly how the direct contribution enters with
different sign on the diagonal and off diagonal Nambu component due
to the loop rule. The way the four point object is connected to the
Green's function is shown in Eq.~\ref{eq:LoopRule-1} and \ref{eq:LoopRule}
for the 21 and 22 element of the self-energy. In the solution of the
gap equation (see Sec.\ \ref{sec:Functional}), this sign difference
will turn out to be crucial in order to have a nontrivial solution
of the gap equation. Note, that the self-energy derived from the Berk-Schrieffer
interaction\cite{PhysRevLett.17.433} satisfies the same sign convention
as derived here.

Under the assumption of singlet SC pairing and magnetic collinearity,
the normal part of the Green's function conserves spin \emph{i.e.}
$G\left(\mathit{1,\!2}\right)=\updelta_{\sigma_{1}\sigma_{2}}G_{\sigma_{1}}\left(\boldsymbol{r}_{1},\!\tau_{1},\!\boldsymbol{r}_{2},\!\tau_{2}\right)$while
the anomalous part flips spin $F\left(\mathit{1,\!2}\right)=\updelta_{\sigma_{1}-\sigma_{2}}F_{\sigma_{1}}\left(\boldsymbol{r}_{1},\!\tau_{1},\!\boldsymbol{r}_{2},\!\tau_{2}\right).$
This aspect has no consequences for the $\bar{\varSigma}^{{\scriptscriptstyle GW}}$
and $\bar{\varSigma}^{{\scriptscriptstyle \text{Ph}}}$ parts of the
self-energy Eqs.~(\ref{eq:SigmaEl}) and (\ref{eq:SigmaPh}), since
the interactions have no spin dependence. However, for the spin-fluctuation
part in Eq.~\ref{eq:SigmaSF-1} the restriction lead to the result
that: 
\begin{align*}
\bar{\varSigma}_{11}\textrm{ and }\bar{\varSigma}_{22} & \textrm{\hspace{10pt} depend only on }\varLambda_{\sigma_{1}\sigma\sigma_{2}\sigma}^{\text{c}{\scriptscriptstyle -}\mathrm{d}}\\
\bar{\varSigma}_{12}\textrm{ and }\bar{\varSigma}_{21} & \textrm{\hspace{10pt} depend only on }\varLambda_{\sigma_{1}\sigma-\sigma\sigma_{2}}^{\text{c}{\scriptscriptstyle +}\text{d}}.
\end{align*}

Furthermore, by comparing with Eq.~\ref{eq:DirectCrossed} it is clear
that $\varLambda^{\text{c}}$ has no $\left(\sigma,\!\sigma,\!-\sigma,\!-\sigma\right)$
component. While $\varLambda^{\text{d}}$ has no component in the
channel $\left(\sigma,\!-\sigma,\!\sigma,\!-\sigma\right)$. Therefore
the following identities hold: 
\begin{align}
\varLambda_{\sigma\sigma-\sigma-\sigma}^{\text{c}{\scriptscriptstyle +}\text{d}}= & \varLambda_{\sigma\sigma-\sigma-\sigma}^{\text{d}}=-\varLambda_{\sigma\sigma-\sigma-\sigma}^{\text{c}{\scriptscriptstyle -}\mathrm{d}}\label{eq:OnlyDirect}\\
\varLambda_{\sigma-\sigma\sigma-\sigma}^{\text{c}{\scriptscriptstyle +}\text{d}}= & \varLambda_{\sigma-\sigma\sigma-\sigma}^{\text{c}}=\varLambda_{\sigma-\sigma\sigma-\sigma}^{\text{c}{\scriptscriptstyle -}\mathrm{d}}\label{eq:OnlyCrossed}
\end{align}
These relations lead to a self-energy containing only $\varLambda^{\text{c-d}}$
and not $\varLambda^{\text{c\ensuremath{\pm}d}}$:

\begin{align}
\bar{\varSigma}_{11}^{{\scriptscriptstyle \text{SF}}} & =-\updelta_{\sigma_{1}\sigma_{2}}G_{\sigma_{1}}\sum_{\sigma}\varLambda_{\sigma_{1}\sigma\sigma_{1}\sigma}^{\text{c}{\scriptscriptstyle -}\mathrm{d}}\label{eq:A1}\\
\bar{\varSigma}_{22}^{{\scriptscriptstyle \text{SF}}} & =\updelta_{\sigma_{1}\sigma_{2}}G_{\sigma_{1}}^{\dagger}\sum_{\sigma}\varLambda_{\sigma\sigma_{1}\sigma\sigma_{1}}^{\text{c}{\scriptscriptstyle -}\mathrm{d}}\label{eq:A2}\\
\bar{\varSigma}_{12}^{{\scriptscriptstyle \text{SF}}} & =\updelta_{\sigma_{1}-\sigma_{2}}F_{\sigma_{1}}\sum_{\sigma}\left(1-2\updelta_{\sigma\sigma_{1}}\right)\varLambda_{\sigma_{1}\sigma-\sigma-\sigma_{1}}^{\text{c}{\scriptscriptstyle -}\mathrm{d}}\label{eq:A3}\\
\bar{\varSigma}_{21}^{{\scriptscriptstyle \text{SF}}} & =-\updelta_{\sigma_{1}-\sigma_{2}}F_{\sigma_{1}}^{\dagger}\sum_{\sigma}\left(1-2\updelta_{\sigma\sigma_{1}}\right)\varLambda_{\sigma\sigma_{1}-\sigma_{1}-\sigma}^{\text{c}{\scriptscriptstyle -}\mathrm{d}}\label{eq:A4}
\end{align}
This is a convenient result, because we have to solve only one BSE
for $\varLambda^{\text{c}{\scriptscriptstyle -}\mathrm{d}}$ (Eq.~\ref{eq:RelevantBSE})
and not the two separate equations for the direct and crossed part.
In the previous expression we use a concise notation in which the
integrals are not written out (compare with Eq.~\ref{eq:SigmaSF-1})
. We will use this notation in the next section when it does not lead
to any ambiguity in the formulae. Unless stated otherwise the coordinates
are contracted analogous to a matrix product.

\subsection{Local Approximation \label{sub:Local-Approximation}}

In the last sections we have constructed an approximate form of the
electronic Nambu self-energy that we believe contains the relevant
contributions to account for a spin-fluctuation mediated pairing.
However, even this approximate form is too complex to be used directly
in simulations on real materials. The dimensionality of the four-point
object $\varLambda^{\text{c}-\text{d}}$ in Eqs.~\ref{eq:A1} to
\ref{eq:A4} and the resulting integrals are simply too complex to
handle. What would make a significant simplification, and bring the
computational cost of the method to an affordable level, would be 
a two-point form of the interaction; meaning an approximate
form that can be written as 
\begin{equation}
\bar{\varSigma}^{{\scriptscriptstyle \text{SF}}}\left(\mathit{1,\!2}\right)_{ab}\overset{!}{=}\bar{G}\left(\mathit{1,\!2}\right)_{ab}\bar{w}^{{\scriptscriptstyle \text{SF}}}\left(\mathit{1,\!2}\right)_{ab},\label{eq:ChangeExact-SigmaSFlocal}
\end{equation}
where $\bar{w}^{{\scriptscriptstyle \text{SF}}}$ is to be understood
as an effective interaction between electrons that accounts for the spin fluctuation 
(SF) pairing and $a,b$ is the index with respect to  the Nambu matrix.
Of course such a form can be obtained by a formal inversion
of the above equation: 
\begin{align*}
\bar{w}^{{\scriptscriptstyle \text{SF}}}\left(\mathit{1,\!2}\right)_{ab} & =\frac{\iint\begin{pmatrix}-G\varLambda^{\text{c}{\scriptscriptstyle -}\mathrm{d}} & F\varLambda^{\text{c}{\scriptscriptstyle +}\text{d}}\\
-{F}^{\dagger}\varLambda^{\text{c}{\scriptscriptstyle +}\text{d}} & {G}^{\dagger}\varLambda^{\text{c}{\scriptscriptstyle -}\mathrm{d}}
\end{pmatrix}_{ab}}{\bar{G}\left(\mathit{1,\!2}\right)_{ab}}
\end{align*}
but this is of no use in practice, because one would need the four-point
object $\varLambda^{\text{c}{\scriptscriptstyle \pm}\text{d}}$ in
the first place.

To obtain a two-point form we make use of an additional approximation,
already common in the context of band structure calculations \citep{ShamKohn_electrongas_PR1966,DelSole_GW_PRB1994,Romaniello_GW_PRB2012},
to use the Kohn-Sham potential as a local approximation for $\varSigma^{{\scriptscriptstyle \text{V}}}$,
namely
\begin{equation}
\varSigma^{{\scriptscriptstyle \text{V}}}\left(\mathit{1,\!2}\right)\approx\updelta_{\tau_1\tau_2}\updelta_{{\bf r}_1{\bf r}_2}v^{\text{xc}}_{\sigma_1\sigma_2}\left({\bf r}_1\tau_1\right).
\end{equation}
The functional derivative (Eq.~\ref{eq:RelevantBSE}) leads to the
xc-kernel $f^{\text{xc}}$, which is a two-point function in space-time
but still a four-point object in spin (${\bf x}_1=\{{\bf r}_1\tau_1\}$): 

\begin{align*}
\frac{\updelta v_{\sigma_{1}\sigma_{2}}^{\text{xc}}\left(\boldsymbol{x}_{1}\right)}{\updelta G(\mathit{3},\!\mathit{4})} & =\sum_{\sigma_{5}\sigma_{6}}\int\!\!\! d\boldsymbol{x}_{5}\underbrace{\frac{\updelta v_{\sigma_{1}\sigma_{2}}^{\text{xc}}\left(\boldsymbol{x}_{1}\right)}{\updelta\rho_{\sigma_{5}\sigma_{6}}(\boldsymbol{x}_{5})}}_{f^{\text{xc}}}\frac{\updelta\rho_{\sigma_{5}\sigma_{6}}(\boldsymbol{x}_{5})}{\updelta G(\mathit{3},\!\mathit{4})}\\
 & =\!\!\iint\!\!\! d\mathit{5}d\mathit{6}f_{\sigma_{1}\sigma_{2}\sigma_{5}\sigma_{6}}^{\text{xc}}\!\left(\boldsymbol{x}_{1}\boldsymbol{x}_{5}\right)\!\frac{\updelta G(\mathit{5},\!6)}{\updelta G(\mathit{3},\!\mathit{4})}\updelta_{\mathit{x}_{5}\mathit{x}_{6}}\\
 & =f_{\sigma_{1}\sigma_{2}\sigma_{3}\sigma_{4}}^{\text{xc}}\left(\boldsymbol{x}_{1}\boldsymbol{x}_{3}\right)\updelta_{\mathit{x}_{3}\mathit{x}_{4}}.
\end{align*}

If Eq.\ \ref{eq:RelevantBSE} is solved with the xc-kernel and the
full $G$ is approximated by the KS one, the well known Dyson equation
from linear response density functional theory appears \citep{PhysRevLett.52.997}:
\begin{align*}
\varLambda^{\text{c}{\scriptscriptstyle -}\mathrm{d}} & =4f^{\text{xc}}+16f^{\text{xc}}\underbrace{G^{{\scriptscriptstyle \text{KS}}}G^{{\scriptscriptstyle \text{KS}}}}_{=\chi^{{\scriptscriptstyle \text{KS}}}}f^{\text{xc}}\\
 & +64f^{\text{xc}}G^{{\scriptscriptstyle \text{KS}}}G^{{\scriptscriptstyle \text{KS}}}f^{\text{xc}}G^{{\scriptscriptstyle \text{KS}}}G^{{\scriptscriptstyle \text{KS}}}f^{\text{xc}}+\dots\\
 & =4f^{\text{xc}}+16f^{\text{xc}}\frac{\chi^{{\scriptscriptstyle \text{KS}}}}{1-f^{\text{xc}}\chi^{{\scriptscriptstyle \text{KS}}}}f^{\text{xc}}
\end{align*}
leading to the proper part of the response function $P_{\sigma_{1}\sigma_{1}\sigma_{2}\sigma_{2}}$.
Since the Green's function is diagonal with respect to spin the longitudinal
and transverse parts of the response decouple: 
\begin{align}
\varLambda_{\sigma_{1}\sigma_{1}\sigma_{2}\sigma_{2}}^{\text{c}{\scriptscriptstyle -}\mathrm{d}}= & 4f_{\sigma_{1}\sigma_{1}\sigma_{2}\sigma_{2}}^{\text{xc}}+\nonumber \\
 & 16\sum_{\sigma_{6}\sigma_{7}}f_{\sigma_{1}\sigma_{1}\sigma_{6}\sigma_{6}}^{\text{xc}}P_{\sigma_{6}\sigma_{6}\sigma_{7}\sigma_{7}}f_{\sigma_{7}\sigma_{7}\sigma_{2}\sigma_{2}}^{\text{xc}}\label{eq:IntDia1}
\end{align}
\begin{align}
\varLambda_{\sigma-\sigma\sigma-\sigma}^{\text{c}{\scriptscriptstyle -}\mathrm{d}} & =4f_{\sigma-\sigma-\sigma\sigma}^{\text{xc}}+\nonumber \\
 & 16f_{\sigma-\sigma-\sigma\sigma}^{\text{xc}}P_{\sigma-\sigma-\sigma\sigma}f_{\sigma-\sigma-\sigma\sigma}^{\text{xc}}.\label{eq:IntFlip1}
\end{align}
The proper part $P$ is related to the full response function $\chi$
via the Dyson equation: 
\begin{align}
\chi_{\sigma_{1}\sigma_{2}\sigma_{3}\sigma_{4}} & =P_{\sigma_{1}\sigma_{2}\sigma_{3}\sigma_{4}}+\nonumber \\
 & \updelta_{\sigma_{1}\sigma_{2}}\delta_{\sigma_{3}\sigma_{4}}\sum_{\sigma\sigma'}P_{\sigma_{1}\sigma_{2}\sigma\sigma}v\chi_{\sigma'\sigma'\sigma_{3}\sigma_{4}}.\label{eq:ProperFullChi}
\end{align}
The response function $\chi_{\sigma_{1}\sigma_{2}\sigma_{3}\sigma_{4}}$
in the spin basis determines the change in the spin resolved charge
density induced by external fields and is defined as: 
\[
\chi_{\sigma_{1}\sigma_{2}\sigma_{3}\sigma_{4}}\left(\boldsymbol{r}_{1},\!\tau_{1},\!\boldsymbol{r}_{2},\!\tau_{2}\right):=\frac{\updelta\rho_{\sigma_{1}\sigma_{2}}\left(\boldsymbol{r}_{1},\!\tau_{1}\right)}{\updelta\varphi_{\sigma_{3}\sigma_{4}}^{\text{ext}}\left(\boldsymbol{r}_{2},\!\tau_{2}\right)}
\]
The equations for $\varLambda^{\text{c-d}}$ will become more transparent
if we rewrite the response quantities on the right hand side of Eqs.\ (\ref{eq:IntDia1})
and (\ref{eq:IntFlip1}) in components of the Pauli matrix \emph{i.e.}:
\begin{align*}
\chi_{ij}\left(\boldsymbol{r}_{1},\!\tau_{1},\!\boldsymbol{r}_{2},\!\tau_{2}\right) & :=\frac{\updelta\rho_{i}\left(\boldsymbol{r}_{1},\!\tau_{1}\right)}{\updelta\varphi_{j}^{\text{ext}}\left(\boldsymbol{r}_{2},\!\tau_{2}\right)}
\end{align*}
In this form $\chi$ represents the change of the electronic charge
$\rho$ or magnetic moment $\boldsymbol{m}$ $\left(\rho_{i}=\left\{ \rho,m_{\mathrm{x}},m_{\mathrm{y},}m_{\mathrm{z}}\right\} \right)$
with respect to physical fields $\left(\varphi_{j}^{\text{ext}}=\left\{ \varphi_{0}^{\text{ext}},B_{\mathrm{x}}^{\text{ext}},B_{\mathrm{y}}^{\text{ext}},B_{\mathrm{z}}^{\text{ext}}\right\} \right)$.
In this work we will label the Pauli index with $i$ and $j$ and
it should not be confused with the Nambu index indicated by $a$ and
$b$ (used in Eq.~\ref{eq:ChangeExact-SigmaSFlocal}). The basis
transformations between the two representations are simply: 
\begin{align*}
A_{\alpha\beta\gamma\delta} & =\frac{1}{4}\sum_{ij}\upsigma_{\alpha\beta}^{i}A_{ij}\upsigma_{\gamma\delta}^{j}\\
A_{ij} & =\sum_{\alpha\beta\gamma\delta}\upsigma_{\beta\alpha}^{i}A_{\alpha\beta\gamma\delta}\upsigma_{\delta\gamma}^{j}
\end{align*}
where $\sigma^{i}$ is the four component vector containing the Pauli
matrices: 
\[
\upsigma^{i}=\left\{ \begin{pmatrix}1 & 0\\
0 & 1
\end{pmatrix},\begin{pmatrix}0 & 1\\
1 & 0
\end{pmatrix},\begin{pmatrix}0 & -\mathrm{i}\\
\mathrm{\mathrm{i}} & 0
\end{pmatrix},\begin{pmatrix}1 & 0\\
0 & -1
\end{pmatrix}\right\} .
\]
Note that the response function is a sparse matrix for the considered
collinear system 
\[
\chi_{ij}=\begin{pmatrix}\chi_{\mathrm{xx}} & \chi_{\mathrm{xy}} & 0 & 0\\
\chi_{\mathrm{yx}} & \chi_{\mathrm{yy}} & 0 & 0\\
0 & 0 & \chi_{\mathrm{zz}} & \chi_{\mathrm{z0}}\\
0 & 0 & \chi_{\mathrm{0z}} & \chi_{00}
\end{pmatrix}
\]
and the proper and full response are equal $\chi_{ij}=P_{ij}$ if
$i,j\in\left\{ \mathrm{x},\mathrm{y}\right\} $ (Eq.\ \ref{eq:ProperFullChi}).
As mentioned above we change the representation of the response function
from spin to the Pauli basis in order to achieve a more transparent
form of the effective interaction: 
\begin{align}
\varLambda_{\sigma_{1}\sigma_{1}\sigma_{2}\sigma_{2}}^{\text{c}{\scriptscriptstyle -}\mathrm{d}} & =\sum_{{\scriptscriptstyle ij\in\{0,z\}}}f_{i\sigma_{1}}^{{\scriptscriptstyle \mathrm{T}}}P_{ij}\left(1-\updelta_{i0}\updelta_{j0}\right)f_{j\sigma_{2}}\label{eq:IntDia}\\
\varLambda_{\sigma-\sigma\sigma-\sigma}^{\text{c}{\scriptscriptstyle -}\mathrm{d}} & =2f_{\sigma}^{{\scriptscriptstyle \text{F}}}\chi_{\sigma}^{{\scriptscriptstyle \text{F}}}f_{\sigma}^{{\scriptscriptstyle \text{F}}}.\label{eq:IntFilp}
\end{align}
where the two point functions $f_{i\sigma}$ and $f_{\sigma}^{{\scriptscriptstyle \text{F}}}$
are given by ($z_{\uparrow}=+1,z_{\downarrow}=-1$): 
\[
f_{z\sigma}^{\mathrm{{\scriptscriptstyle \mathrm{T}}}}:=z_{\sigma}f_{\mathrm{zz}}^{\text{xc}}+f_{\mathrm{0z}}^{\text{xc}}\ \ \ f_{\mathrm{z\sigma}}:=z_{\sigma}f_{\mathrm{zz}}^{\text{xc}}+f_{\mathrm{z0}}^{\text{xc}}
\]
\[
f_{0\sigma}^{\mathrm{{\scriptscriptstyle \mathrm{T}}}}:=f_{00}^{\text{xc}}+z_{\sigma}f_{\mathrm{z0}}^{\text{xc}}\ \ \ f_{0\sigma}:=f_{00}^{\text{xc}}+z_{\sigma}f_{0z}^{\text{xc}}
\]
\[
f_{\sigma}^{{\scriptscriptstyle \text{F}}}:=f_{\mathrm{xx}}^{\text{xc}}+z_{\sigma}\mathrm{i}f_{\mathrm{xy}}^{\text{xc}}\ \ \ \chi_{\sigma}^{{\scriptscriptstyle \text{F}}}:=\chi_{\mathrm{xx}}+z_{\sigma}i\chi_{\mathrm{xy}}.
\]
In Eq.~\ref{eq:IntDia} we have dropped $f_{\sigma_{1}\sigma_{1}\sigma_{2}\sigma_{2}}^{\text{xc}}+f_{0\sigma_{1}}^{\mathrm{T}}P_{00}f_{0\sigma_{2}}$
in order to avoid any double counting: This term is in fact already
accounted for by the screened Coulomb interaction $w$ in the $GW$
term, that contains an analogous contribution in the form $v+vP_{00}v+\dots\ .$
In addition we neglect the linear order $f_{\sigma-\sigma-\sigma\sigma}^{\text{xc}}$,
because in a system featuring magnetic fluctuations it is supposed
to be small as compared to the dominant $f_{\sigma}^{{\scriptscriptstyle \text{F}}}\chi_{\sigma}^{{\scriptscriptstyle \text{F}}}f_{\sigma}^{{\scriptscriptstyle \text{F}}}$
term. This is because the spin-fluctuations should appear as a large
value of the magnetic susceptibility.

The form of $\varLambda^{\text{c}{\scriptscriptstyle -}\mathrm{d}}$
in Eq.~\ref{eq:IntFilp} has now obtained an immediate physical interpretation:
The exchange-correlation kernels $f^{\text{xc}}$ act as a vertex
for the electronic interaction mediated by spin-fluctuations, which
are expressed by the magnetic susceptibility $\chi$.

The transverse part allows for a flip of the electronic spin, which
can be understood in the following way: 
\begin{enumerate}
\item The spin-flip of electron 1 corresponds to a local fluctuation in
the magnetic moment $\updelta m_{1}$. 
\item This in turn creates a magnetic field via the kernel: $\updelta B_{1}=f^{\text{xc}}\updelta m_{1}$. 
\item If the system features magnetic fluctuations the $\updelta B_{1}$
leads to fluctuations in the system: $\updelta m_{2}=\chi\updelta B_{1}$. 
\item The fluctuation (magnons) couple via the second kernel to another
electron $\updelta B_{2}=f^{\text{xc}}\updelta m_{2}$, whose spin
is flipped in the absorption process. 
\end{enumerate}
This interpretation is analogous to the one given by Kukkonen and
Overhauser for the charge fluctuations \cite{PhysRevB.20.550} and
shows that the term $\varLambda^{\text{c}{\scriptscriptstyle -}\mathrm{d}}$
in the local form represents an effective interaction between electrons
mediated by magnetic fluctuations.

The final $\bar{\varSigma}^{{\scriptscriptstyle \text{SF}}}$ is constructed
by inserting the two-point particle-hole propagators given in Eqs.\ \ref{eq:IntDia}
and \ref{eq:IntFilp} in the equation for the self-energy Eqs.\ \ref{eq:A1}
to \ref{eq:A4}. We do this in the next section. Note that by doing
so, a separation in direct and crossed contribution is implied for
the xc-kernel (see Eqs.~\ref{eq:OnlyDirect} and~\ref{eq:OnlyCrossed}).
This is an assumption because the xc-kernel is in general not based
on a diagrammatic expansion.

\section{Final Form of the Self-Energy\label{sec:FinalForm}}

So far our formalism has been derived for collinear magnetic systems.
We will now simplify it for the case of a non-magnetic system. This
means that, by construction, we will not consider the possibility
of atomic scale coexistence between magnetism and SC. We believe that
this assumption is justified for a large set of high-temperature SC
(cuprates and pnictides) where usually (although exceptions have been
observed) the antiferromagnetic (AFM) order is completely suppressed
in the SC regime\citep{Nature.464.183,RevModPhys.78.17}.

In a non magnetic system the response functions and xc-kernel are
diagonal with respect to the Pauli index and the three directions
with respect to the magnetic field are degenerate. In this case the
effective interaction in Eqs.~\ref{eq:IntDia} and~\ref{eq:IntFilp}
reduces to the a simple form (we use here $\boldsymbol{x}$ is a combined
variable of space and time $\boldsymbol{x}=\left\{ \boldsymbol{r}\tau\right\} $):
\begin{align*}
\varLambda_{\sigma_{1}\sigma_{1}\sigma_{2}\sigma_{2}}^{\text{c}{\scriptscriptstyle -}\mathrm{d}}\left(\boldsymbol{x}_{1},\!\boldsymbol{x}_{2}\right) & =z_{\sigma_1}z_{\sigma_2}\frac{1}{2}\varLambda^{{\scriptscriptstyle \text{SF}}}\left(\boldsymbol{x}_{1},\!\boldsymbol{x}_{2}\right)\\
\varLambda_{\sigma-\sigma\sigma-\sigma}^{\text{c}{\scriptscriptstyle -}\mathrm{d}}\left(\boldsymbol{x}_{1},\!\boldsymbol{x}_{2}\right) & =\varLambda^{{\scriptscriptstyle \text{SF}}}\left(\boldsymbol{x}_{1},\!\boldsymbol{x}_{2}\right)
\end{align*}
\begin{align}
\varLambda^{{\scriptscriptstyle \text{SF}}}\left(\boldsymbol{x}_{1},\!\boldsymbol{x}_{2}\right) & :=2\!\iint\!\!\mathrm{d}\boldsymbol{x}\mathrm{d}\boldsymbol{x}'\times\nonumber \\
 & f_{zz}^{\text{xc}}\left(\boldsymbol{x}_{1},\!\boldsymbol{x}_{2}\right)\chi_{zz}\left(\boldsymbol{x}\boldsymbol{x}'\right)f_{zz}^{\text{xc}}\left(\boldsymbol{x}_{1},\!\boldsymbol{x}_{2}\right)\label{eq:InteractionSF}
\end{align}
and we insert this form in Eqs.\ (\ref{eq:A1}) to (\ref{eq:A4}):
\begin{align}
\bar{\varSigma}_{ab}^{{\scriptscriptstyle \text{SF}}}\left(\boldsymbol{x}_{1},\!\boldsymbol{x}_{2}\right) & =\frac{3}{2}\left(-1\right)^{b+1}\varLambda^{{\scriptscriptstyle \text{SF}}}\left(\boldsymbol{x}_{1},\!\boldsymbol{x}_{2}\right)\bar{G}_{ab}\left(\boldsymbol{x}_{1},\!\boldsymbol{x}_{2}\right).\label{eq:SigmaSF}
\end{align}
The prefactor represents the fact, that the diagonal part
enters with the opposite sign due to the effect of the loop rule discussed
in Sec.~\ref{sub:InvestigationParticleHole}. By construction the
equation has the $GW$ form, however with an interaction originating from
 spin-fluctuation and denoted as $\varLambda^{{\scriptscriptstyle \text{SF}}}$.
Note that this effective interaction, in the limit of an homogeneous electron gas, reduces to the form derived by Vignale and Singwi in Ref.~\onlinecite{PhysRevB.32.2156}.
$\varLambda^{{\scriptscriptstyle \text{SF}}}$ contains the xc-kernel
and the magnetic response function, which can be calculated using
TD-DFT\cite{PhysRevLett.52.997}. The total self-energy is given by
the sum of $\bar{\varSigma}^{{\scriptscriptstyle GW}},\bar{\varSigma}^{{\scriptscriptstyle \text{SF}}}$
and $\bar{\varSigma}^{{\scriptscriptstyle \text{Ph}}}$ given in Eqs.\ \ref{eq:SigmaEl},
\ref{eq:SigmaSF} and \ref{eq:SigmaPh}, respectively.

\section{The Functional\label{sec:Functional}}

So far we have derived the contribution from the spin-fluctuations
to the self-energy, and correspondingly a spin-fluctuation pairing
that can be used in any theory of SC. In this section we will specialize
this result to be used within the framework of SCDFT. To do this we
will make use of the Sham-Schlüter connection\citep{PhysRevLett.51.1888}
between a KS and an interacting system, generalized to the SC case
by Marques\citep{PhDMarques}. We will assume that $v^{\text{xc}}$ 
and the  diagonal part of $\bar{\varSigma}^{{\scriptscriptstyle GW}}$
act in a similar way as a mass operator on the Hartree states and
cancel each other. Then the non-interacting SC-KS is mapped to the
interacting system by the following self-energy form: 
\begin{equation}
\bar{\varSigma}^{\ensuremath{{\scriptscriptstyle \mathrm{SS}}}}=\bar{\varSigma}^{{\scriptscriptstyle GW}}+\bar{\varSigma}^{{\scriptscriptstyle \text{SF}}}+\bar{\varSigma}^{{\scriptscriptstyle \text{Ph}}}-\begin{pmatrix}GW & {\varDelta^{\text{xc}}}^{*}\\
\varDelta^{\text{xc}} & -{G}^{\dagger}w
\end{pmatrix}.
\end{equation}
The Sham-Schlüter connection follows by imposing that the total density
$\rho\left(\boldsymbol{r}_{1}\right)=\lim_{\boldsymbol{r}_{1}\rightarrow\boldsymbol{r}_{2}}\frac{2}{\beta}\sum_{\omega_{n}}G\left(\boldsymbol{r}_{1},\!\boldsymbol{r}_{2},\!\omega_{n}\right)$
and the anomalous density $\chi\left(\boldsymbol{r}_{1},\!\boldsymbol{r}_{2}\right)=\sum_{\omega_{n}}F\left(\boldsymbol{r}_{1},\!\boldsymbol{r}_{2},\!\omega_{n}\right)$
are identical in the KS and interacting system: 
\begin{align*}
0 & =\updelta_{ab}\lim_{\boldsymbol{r}_{1}\rightarrow\boldsymbol{r}_{2}}\frac{2}{\beta}\sum_{\omega_{n}}\mathrm{e}^{\mathrm{i}\omega_{n}0^{+}}\left[\bar{G}^{{\scriptscriptstyle \text{KS}}}\bar{\varSigma}^{\ensuremath{{\scriptscriptstyle \mathrm{SS}}}}\bar{G}\right]_{ab}\\
0 & =\left(1-\updelta_{ab}\right)\frac{1}{\beta}\sum_{\omega_{n}}\mathrm{e}^{\mathrm{i}\omega_{n}0^{+}}\left[\bar{G}^{{\scriptscriptstyle \text{KS}}}\bar{\varSigma}^{\ensuremath{{\scriptscriptstyle \mathrm{SS}}}}\bar{G}\right]_{ab}.
\end{align*}
The connection becomes a closed equation for the superconducting gap
by approximating the full Green's function on the right hand side
and the one in $\bar{\varSigma}$ with the KS one. In addition we neglect
all contributions that are explicitly higher than linear in the pairing
potential. Since, as discussed in Sec.~\ref{sec:Extension_of_Sigma},
we are mostly concerned with computing an accurate critical temperature,
rather than the full temperature dependence of the superconducting gap.
In this approximation, the $12$ element of the Sham-Schlüter equation
simplifies to: 
\begin{align*}
0 & =\frac{1}{\beta}\sum_{n}\mathrm{e}^{\mathrm{i}\omega_{n}0^{+}}G^{{\scriptscriptstyle \text{KS}}}{G^{{\scriptscriptstyle \text{KS}}}}^{\dagger}{\varDelta^{\text{xc}}}^{*}\\
 & +\frac{1}{\beta}\sum_{n}\mathrm{e}^{\mathrm{i}\omega_{n}0^{+}}G^{{\scriptscriptstyle \text{KS}}}\bar{\varSigma}_{11}F^{{\scriptscriptstyle \text{KS}}}\\
 & -\frac{1}{\beta}\sum_{n}\mathrm{e}^{\mathrm{i}\omega_{n}0^{+}}G^{{\scriptscriptstyle \text{KS}}}\bar{\varSigma}_{12}{G^{{\scriptscriptstyle \text{KS}}}}^{\dagger}.
\end{align*}
The Matsubara summation may be evaluated analytically because the
frequency dependence of the KS Green's function is known and for the
response functions and phonons a frequency representation with respect
to the anti-Hermitian part of the retarded quantities holds (Eq.~\ref{eq:KKrelation}).
The evaluation is done with the help of the residue theorem, which,
 for the Matsubara summation of an analytic function $A\left(z\right)$, leads
to: 
\begin{equation}
\frac{1}{\beta}\sum_{n}^{\infty}A\left(\mathrm{i}\omega_{n}\right)=\sum_{m}^{\text{Poles}\in\gamma}\text{res}\left[f_{\beta}\left(z\right)A\left(z\right),z_{m}\right],
\end{equation}
where the contour $\gamma$ are two infinite half-circle excluding
the imaginary axis and $f_{\beta}$ is the Fermi distribution function.
At this point an adiabatic approximation for the xc-kernel is assumed.
This reduces the order of poles in $\varLambda^{{\scriptscriptstyle \text{SF}}}$
and the same residue are found for the Coulomb, Phonon and spin-fluctuation
contribution. After the evaluation of the Matsubara sum\cite{PhysRevB.72.024545},
the equation is inverted for $\varDelta_{k}^{\text{xc}},$ leading
to a gap equation very similar to the conventional one in Eq.~\ref{eq:gap}.

\begin{widetext} 
\begin{align}
\varDelta_{k}^{\text{xc}} & =-\varDelta_{k}^{\text{xc}}\mathcal{Z}_{k}^{{\scriptscriptstyle D}}-\sum_{k'}\mathcal{K}_{kk'}^{{\scriptscriptstyle C}}\frac{\tanh\left(\frac{\beta E_{k'}}{2}\right)}{2E_{k'}}\varDelta_{k'}^{\text{xc}}\label{eq:GapFinal}
\end{align}
\begin{align*}
\mathcal{Z}_{k}^{{\scriptscriptstyle D}} & =\frac{1}{\uppi}\sum_{k'}\int_{0}^{\infty}\mathrm{d}\omega\frac{\mathfrak{Im}\left[\frac{3}{2}\varLambda_{kk'}^{{\scriptscriptstyle \text{SF}}}\left(\omega\right)\right]-\varLambda_{kk'}^{{\scriptscriptstyle \text{Ph}}}\left(\omega\right)}{2\tanh\left(\frac{\beta\zeta_{k}}{2}\right)}\frac{d}{d\zeta_{k}}J^{+}\left(\zeta_{k},\zeta_{k'}\omega\right)
\end{align*}
\[
\mathcal{K}_{kk'}^{{\scriptscriptstyle C}}=\frac{2}{\uppi}\int_{0}^{\infty}\mathrm{d}\omega\frac{\mathfrak{Im}\left[w_{kk'}\left(\omega\right)+\frac{3}{2}\varLambda_{kk'}^{{\scriptscriptstyle \text{SF}}}\left(\omega\right)\right]+\varLambda_{kk'}^{{\scriptscriptstyle \text{Ph}}}\left(\omega\right)}{\tanh\left(\frac{\beta\zeta_{k}}{2}\right)\tanh\left(\frac{\beta\zeta_{k'}}{2}\right)}J^{-}\left(\zeta_{k},\zeta_{k'}\omega\right)
\]
\begin{align*}
I_{\beta}\left(E_{k},\! E_{k'},\!\omega\right) & :=-\mathrm{f}_{\beta}\left(E_{k}\right)\mathrm{f}_{\beta}\left(E_{k'}\right)\mathrm{b}_{\beta}\left(\omega\right)\left[\frac{\mathrm{e}^{\beta E_{k}}-\mathrm{e}^{\beta\left(E_{k'}+\omega\right)}}{E_{k}-E_{k'}-\omega}-\frac{\mathrm{e}^{\beta E_{k'}}-\mathrm{e}^{\beta\left(E_{k}+\omega\right)}}{E_{k}-E_{k'}+\omega}\right]\\
J_{\beta}^{\pm}\left(\zeta_{k},\!\zeta_{k'},\!\omega\right) & :=I_{\beta}\left(\zeta_{k},\zeta_{k'}\omega\right)\pm I_{\beta}\left(\zeta_{k},-\zeta_{k'}\omega\right).
\end{align*}

\end{widetext}

$\mathrm{b}_{\beta}\left(\omega\right)$ is the Bose distribution
function and $\zeta_{k}$ are the single particle KS energies of the
non-SC system relative to the chemical potential $\zeta_{k}=\epsilon_{k}-\mu$.
The kernels in this integral equation represent different physical
processes introduced by the corresponding self-energy contribution. 
\begin{enumerate}
\item The $\varLambda_{kk'}^{{\scriptscriptstyle \text{Ph}}}\left(\omega\right)$
term describes pairing between electrons due to phonons. The interaction
is attractive: $\varLambda_{kk'}^{{\scriptscriptstyle \text{Ph}}}\left(\omega\right)<0$. 
\item The $w_{kk'}\left(\omega\right)$ term is the scattering of electrons
due to Coulomb interaction. The bare Coulomb interaction is reduced
by intermediate scattering processes (screening) $w=v\epsilon^{{\scriptscriptstyle -1}}$
(Eq.~\ref{eq:dielectric}). Plasmonic effects may also enter via
this term. 
\item The last term $\varLambda_{kk'}^{{\scriptscriptstyle \text{SF}}}\left(\omega\right)$
contains the magnetic response function $\chi_{zz}$ and hence becomes
important if the system is close to a transition to a magnetic phase.
In such a case the response function features sharp excitations, which
represent paramagnons. 
\end{enumerate}
The two last terms originate both from the Coulomb interaction and
are therefore intrinsically repulsive: 
\begin{equation}
\mathfrak{Im}\left[\varLambda_{kk'}^{{\scriptscriptstyle \text{SF}}}\left(\omega\right)\right]>0\text{ and }\mathfrak{Im}\left[w_{kk'}\left(\omega\right)\right]>0.
\end{equation}
If these are the strongest terms in the gap equation~\ref{eq:GapFinal},
in order to have a non-trivial solution, a sign change must occur
in the gap function. We will show how this mechanism works in the
following section where we apply the formalism to a model system and
investigate the general structure of the theory.

\section{Application to a Two Band Model System\label{sec:TwoBandModel}}

\subsection{Isotropic approximation and the two band model with a SF pairing}

The function $\varDelta_{n\boldsymbol{k}}$ is known to have a strong
dependence on the $\epsilon_{n\boldsymbol{k}}$\cite{MarquesSCDFTII2005}.
The remaining $\boldsymbol{k}$-space structure however is often of
little importance, especially within topologically connected Fermi
surface portions \cite{Floris_Pb_PRB2007}. Therefore it is convenient
to define an isotropic (or multi-band-isotropic) approximation, by
means of the following averaging operation: 
\begin{align*}
\varDelta_{n\boldsymbol{k}} & \approx\varDelta_{n}\left(E\right):=\frac{1}{N_{n}\left(E\right)}\sum_{\boldsymbol{k}}\updelta\left(\epsilon_{n\boldsymbol{k}}-E\right)\varDelta_{n\boldsymbol{k}}\\
N_{n}\left(E\right) & :=\sum_{\boldsymbol{k}}\updelta\left(\epsilon_{n\boldsymbol{k}}-E\right).
\end{align*}
where $N_{n}\left(E\right)$ is the density of states of band $n$.
This simplification leads to an isotropic gap equation, where all
interactions are replaced by energy and band dependent quantities.
As an example of how this averaging works, we consider the spin-fluctuation
term: 
\begin{align*}
\varLambda_{nn'}^{{\scriptscriptstyle \text{SF}}}\left(E,\! E',\!\omega\right) & \approx\frac{1}{N_{n}\left(E\right)}\sum_{\boldsymbol{k}\boldsymbol{k}'}\varLambda_{n\boldsymbol{k}n'\boldsymbol{k}'}^{{\scriptscriptstyle \text{SF}}}\left(\omega\right)\times\\
 & \updelta\left(\epsilon_{n\boldsymbol{k}}-E\right)\updelta\left(\epsilon_{n'\boldsymbol{k}'}-E'\right).
\end{align*}
We assume the system to be close to an AFM instability with the ordering
and nesting vector $\boldsymbol{q}_{\text{c}}$. Then the proximity
to the magnetic phase leads to strong fluctuations (paramagnons) at
low frequencies and the vector $\boldsymbol{q}_{\text{c}}$ in the
magnetic response function $\chi_{zz}\left(\omega,\!\boldsymbol{q}\right)$
\cite{PhysRevB.86.060412}. These fluctuations are expected to be
weak for other vectors. The portions (bands) of the Fermi surface
nested by $\boldsymbol{q}_{\text{c}}$ will be labeled as $n=+$ and
$n=-$.

The usual TDDFT kernels like the adiabatic local density approximation
have no dependence on $\left(\omega,\!\boldsymbol{q}\right)$ and
the form of $\varLambda^{{\scriptscriptstyle \text{SF}}}$ in frequency
and $\boldsymbol{q}$ is determined by $\chi_{zz}$.

In such a situation the isotropic effective interaction is expected
to be small for intra-band scattering $\left(\varLambda_{\pm\pm}^{{\scriptscriptstyle \text{SF}}}\approx0\right)$
and peaked for inter-band scattering $\left(\varLambda_{\pm\mp}^{{\scriptscriptstyle \text{SF}}}\right)$.

This situation is modeled by a simple parabola centered around a characteristic
frequency $\bar{\omega}$ (see Fig.~\ref{fig:GapOfKappa} top right
- we have also tested a Gaussian and a Lorentzian form and we find
that the shape has little effect on the properties of the model):

\begin{equation}
\varLambda_{{\scriptscriptstyle IJ}}^{{\scriptscriptstyle \text{SF}}}\left(E,\! E',\!\omega\right)=\begin{cases}
c_{1}N_{{\scriptscriptstyle J}}\left(E'\right)\left[1-\left(\frac{\omega-\bar{\omega}-\frac{c_{2}}{2}}{c_{2}}\right)^{2}\right]\\
\text{if }\left|\omega-\bar{\omega}\right|\leq\frac{c_{2}}{2}\text{ and }I\neq J\\
\\
0\text{ elsewhere.}
\end{cases}\label{eq:ModelLambda}
\end{equation}

We fix the width $c_{2}$ to a value of $0.01\text{ Ryd}$ and the
density of states times the peak height $c_{1}$ is determined by
requiring a value for the effective coupling strength $\lambda^{{\scriptscriptstyle \text{SF}}}$.
The effective coupling strength is given by the integral of $\varLambda_{{\scriptscriptstyle IJ}}^{{\scriptscriptstyle \text{SF}}}$
with respect to $\omega$: 
\[
\lambda_{{\scriptscriptstyle IJ}}^{{\scriptscriptstyle \text{SF}}}:=\frac{3}{2\uppi}\int_{0}^{\infty}\mathrm{d}\omega\frac{2\varLambda_{{\scriptscriptstyle IJ}}^{{\scriptscriptstyle \text{SF}}}\left(\epsilon_{{\scriptscriptstyle \text{F}}},\!\epsilon_{{\scriptscriptstyle \text{F}}},\!\omega\right)}{\omega}.
\]
Within this simple multiband isotropic model the structure of our
spin-fluctuation theory of SC can be made more transparent.

\subsection{Discussion of the SF Contribution}

Here and in the next section, we will assume a two-band isotropic
approximation discussed in the previous section. In this way we try to
investigate the general solution of the SCDFT gap equation for a SF
mediated pairing. As a first step we neglect completely Coulomb and
phonon contributions, considering only the SF interaction given in
Eq.~\ref{eq:ModelLambda}.

\begin{figure}[h]
\begin{centering}
\includegraphics[width=0.95\columnwidth]{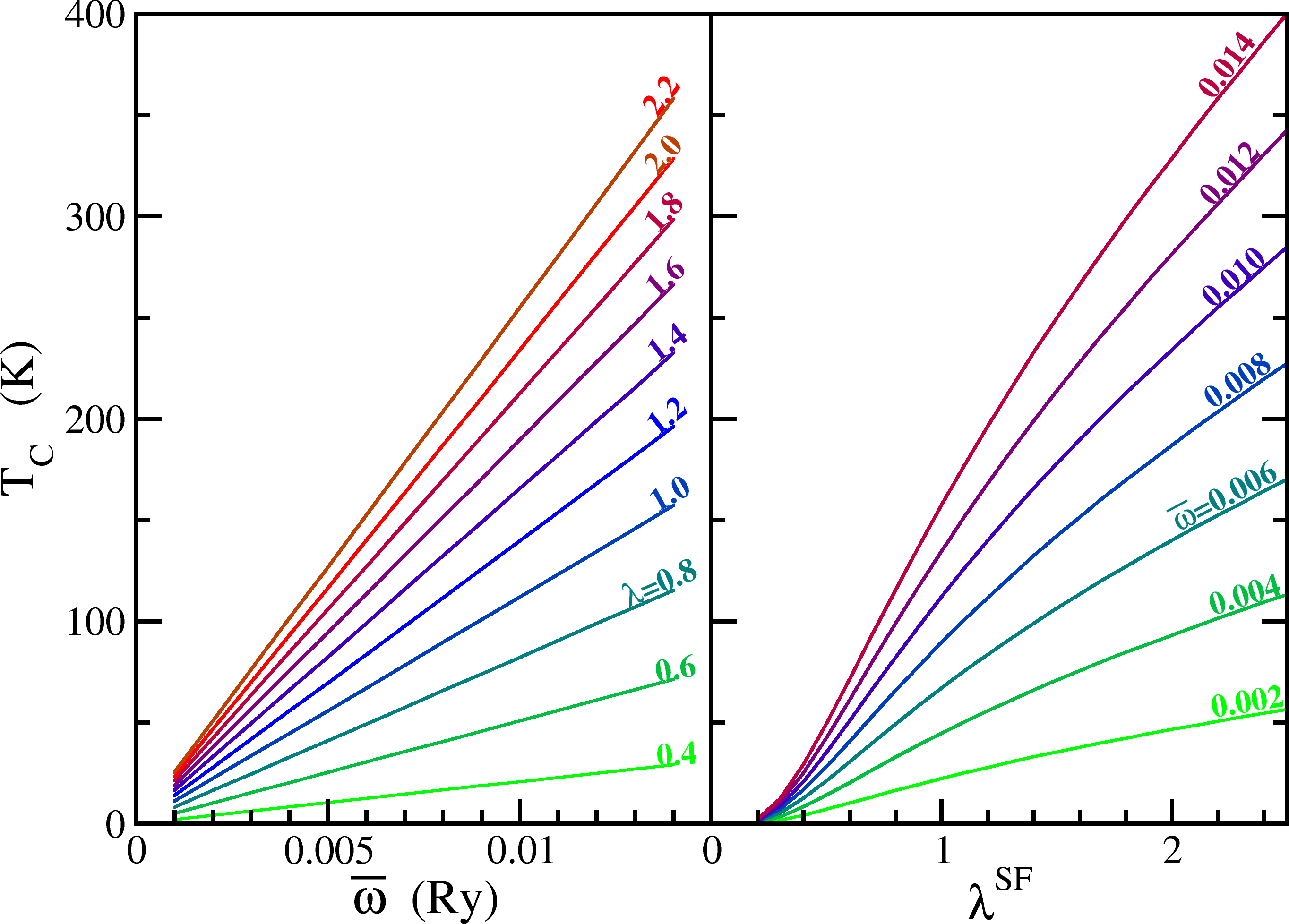} 
\par\end{centering}

\caption{(left) $T_{\text{c}}$ for different $\lambda^{{\scriptscriptstyle \text{SF}}}$
as a function of the average $\bar{\omega}$. (right) Analogous plot,
here for several $\bar{\omega}$ as a function of $\lambda^{{\scriptscriptstyle \text{SF}}}$.\label{fig:SolutionOverView}}
\end{figure}

We modify the SF by acting on the parameters $\bar{\omega}$ and $\lambda_{{\scriptscriptstyle IJ}}^{{\scriptscriptstyle \text{SF}}}$.
In Fig.~\ref{fig:SolutionOverView} we show the critical temperature
as a function of $\bar{\omega}$ and $\lambda_{{\scriptscriptstyle IJ}}^{{\scriptscriptstyle \text{SF}}}$.
From Eliashberg theory for phonon driven superconductors we have knowledge
of the following relations between characteristic frequency and average
coupling strength\cite{PhysRev.167.331,PhysRevB.12.905}:

\begin{align}
T_{\text{c}} & \propto\bar{\omega}\,\mathrm{e}^{-\alpha/\lambda}\text{ for small }\lambda\label{eq:ExpTc}\\
T_{\text{c}} & \propto\bar{\omega}\,\sqrt{\lambda}\text{ for large }\lambda.\label{eq:SqrtTc}
\end{align}
On the right hand side of Fig.~\ref{fig:SolutionOverView} we can
recognize the exponential and square root behavior with respect to
$\lambda^{{\scriptscriptstyle \text{SF}}}$. The transition between
small and large coupling takes place at $\lambda^{{\scriptscriptstyle \text{SF}}}\sim1.5$
for $\bar{\omega}=0.15$~Ryd.
For the dependence of $T_{\text{c }}$ with respect to $\bar{\omega}$,
we find a linear behavior.

This result is not accidental, because the sign change of the gap
leads effectively to an attractive interaction between the two bands,
therefore within this simplified model there is no formal difference
between spin-fluctuation repulsive pairing and conventional phononic
attraction.

Within such a model calculation we can estimate the coupling strength
in the iron based superconductors, simply by the experimental knowledge
that the characteristic energy of the magnetic fluctuations are of
about $20\text{ meV}$ \cite{RevModPhys.83.1589}. This implies a
coupling $\lambda^{{\scriptscriptstyle \text{SF}}}$ of about 1 (neglecting
phononic and Coulomb effects) to reach the critical temperatures of
$\sim10\text{ K}$ found in these compounds.

\subsection{Interplay between Coulomb, Spin-Fluctuation and Phonon Contribution}

In the previous section we have observed that the features of the
SCDFT gap equation with a SF interaction is relatively simple and
similar to the conventional phononic case. Here we add the effect
of phonon and Coulomb interactions. This will create a frustration
on the SC potential because the three interaction will compete against
each other.

We use the same spin-fluctuation spectrum in Eq.\ (\ref{eq:ModelLambda})
and fix $\lambda^{{\scriptscriptstyle \text{SF}}}=1.2$ and $\bar{\omega}=0.01\text{ Ryd}$.
The Coulomb interaction is very different in nature, compared to the
SF. In particular its frequency dependence develops in the plasmonic
energy scale (eV). We therefore ignore it, and assume a flat interaction
with respect to $\omega$.

\begin{figure}[h]
\begin{centering}
\includegraphics[width=0.95\columnwidth]{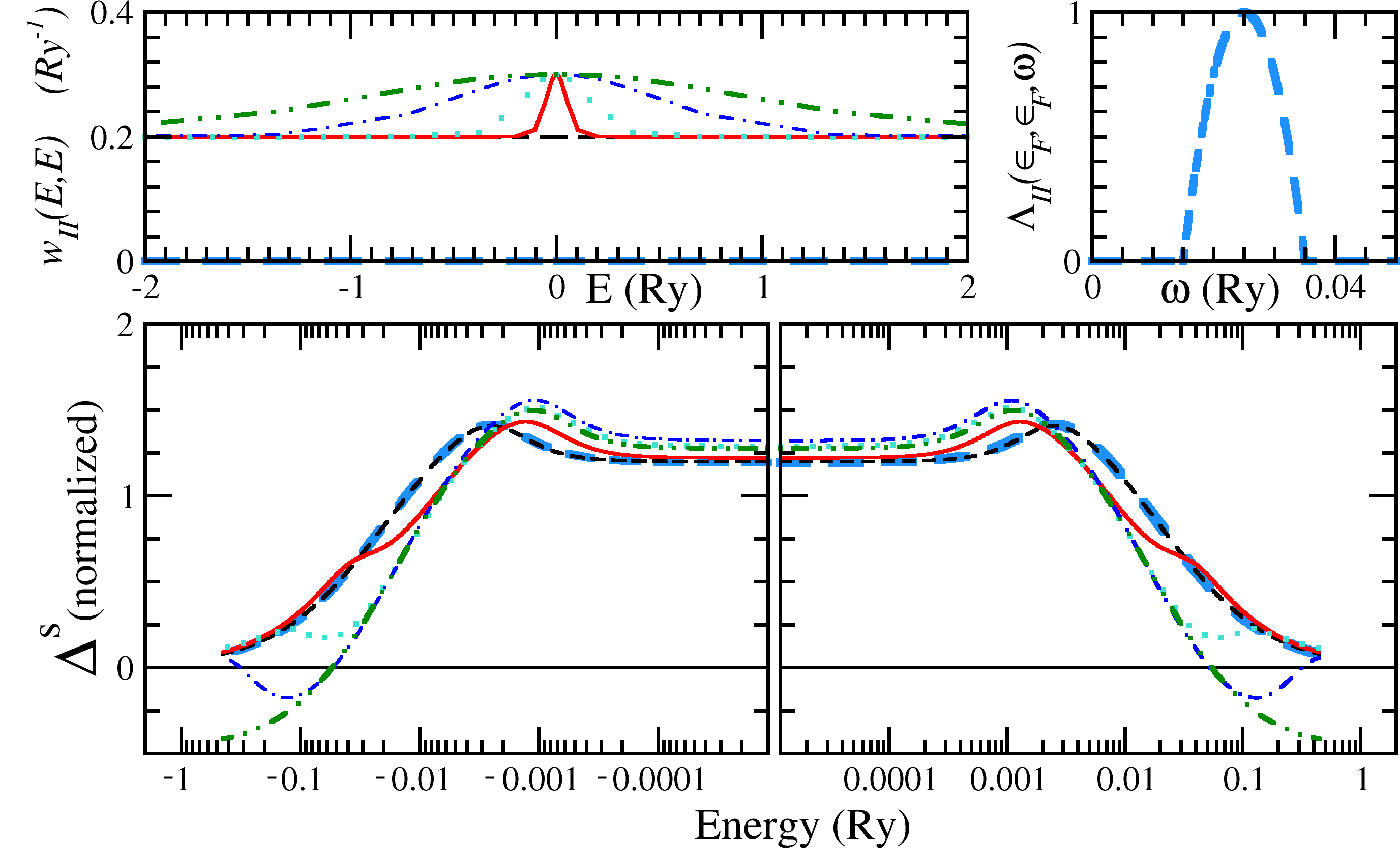} 
\par\end{centering}

\caption{(top-left) Form of the Coulomb interaction for various $\kappa$ determining
the decay of the Coulomb interaction (Eq.\ \ref{eq:ConstructFxc-ModelCoulomb}).
(top-right) Form of the SF interaction Eq.\ \ref{eq:ModelLambda}.
(bottom) Gap function $\varDelta_{+}$ as a function of energy and
$\kappa$. \label{fig:GapOfKappa}}
\end{figure}

It is expected, that this interaction decays like $\frac{1}{\boldsymbol{q}^{2}+\boldsymbol{q}_{\text{TF}}^{2}}$,
where $\boldsymbol{q}_{\text{TF}}$ is the Thomas-Fermi screening
vector. Therefore the contribution for small momentum transfer (intra-band)
should be much larger than the inter-band contribution corresponding
to $\boldsymbol{q}_{\text{c}}\approx\boldsymbol{k}-\boldsymbol{k}'$.
Similarly for scattering from the Fermi level to high-energy states,
the scattering should become momentum independent\cite{Massidda_Sanna_SUST}.
We model this picture in the following way: 
\begin{equation}
w_{{\scriptscriptstyle IJ}}\left(E_{1},\! E_{2}\right)=\begin{cases}
N_{{\scriptscriptstyle J}}\left(E_{2}\right)\left(U_{0}+U_{1}\mathrm{e}^{-\kappa\left(E_{1}^{2}+E{}_{2}^{2}\right)}\right)\\
\text{if }I=J\\
\\
N_{{\scriptscriptstyle J}}\left(E_{2}\right)U_{0}\\
\text{if }I\neq J.
\end{cases}\label{eq:ConstructFxc-ModelCoulomb}
\end{equation}
The diagonal part $w_{{\scriptscriptstyle II}}\left(E_{1},\! E_{1}\right)$
of this interaction is shown in Fig.~\ref{fig:GapOfKappa}. For the
parameters of the Coulomb interaction we choose $N_{{\scriptscriptstyle J}}\left(\epsilon_{{\scriptscriptstyle \text{F}}}\right)U_{0}=0.2,$
$U_{1}=\frac{U_{0}}{2}$. The parameter $\kappa$ is used to control
the Coulomb interaction: If $\kappa$ is large the Coulomb interaction
decays very quickly in energy.

Due to the choice of an electron hole symmetric DOS ($N_{+}\left(E\right)=N_{-}\left(E\right)$)
and interaction, the gap function is also totally symmetric: $\varDelta_{{\scriptscriptstyle I}}\left(E\right)=\varDelta_{{\scriptscriptstyle I}}\left(-E\right)$
and $\varDelta_{+}\left(E\right)=-\varDelta_{-}\left(E\right)$ and
hence only the positive branch $\varDelta_{+}\left(E\right)$ is shown
in Fig.~\ref{fig:GapOfKappa}. Note that within this symmetry a constant
coulomb interaction ($\kappa\to\infty$) cancels out completely from
the gap equation~\ref{eq:GapFinal}. 

In general the gap function shows a typical form, being constant close
to the Fermi level, followed by an extremum and a decay for larger
energies \cite{PhDSanna,MarquesSCDFTII2005}. By decreasing the value
of $\kappa$ the Coulomb contribution starts to influence the results.
The critical temperature decreases, due to repulsion within one band
and the gap starts to show dips. The dips indicate the regime, where
the Coulomb interaction competes with the spin-fluctuation. For $\kappa<1$
the Coulomb contribution are strong enough to flip the sign of the
gap function for certain energies. The sign change of the gap function
at higher energies reduces the effect related to the repulsive Coulomb
term in the gap equation~\ref{eq:GapFinal}.

Effectively, the Coulomb contribution on the full energy scale may
be mapped to a reduced effective Coulomb term on a smaller energy
scale due to the sign change of the gap function. Hence, the sign
change of the gap function is the way Coulomb renormalization happens
in SCDFT\cite{MarquesSCDFTII2005,Massidda_Sanna_SUST}. Note that
the sign change of the gap happens far away from the Fermi level.

However, for $\kappa=4$ the Coulomb contribution still decays faster
in energy than the spin-fluctuation term, which leads to one more
sign change in the large energy regime (dash-dotted blue line in Fig.\ \ref{fig:GapOfKappa}).
If we decrease the $\kappa$ further the Coulomb contribution dominate
also in the large energy range and the gap changes sign only once.

Note, that the critical temperature converges quickly with respect
to $\kappa.$ This indicates, that the Coulomb interaction influences
the critical temperature only in a small energy window for the symmetric
two band system and the renormalization of the gap is not effecting
the critical temperature strongly.

\begin{figure}[h]
\begin{centering}
\includegraphics[width=0.95\columnwidth]{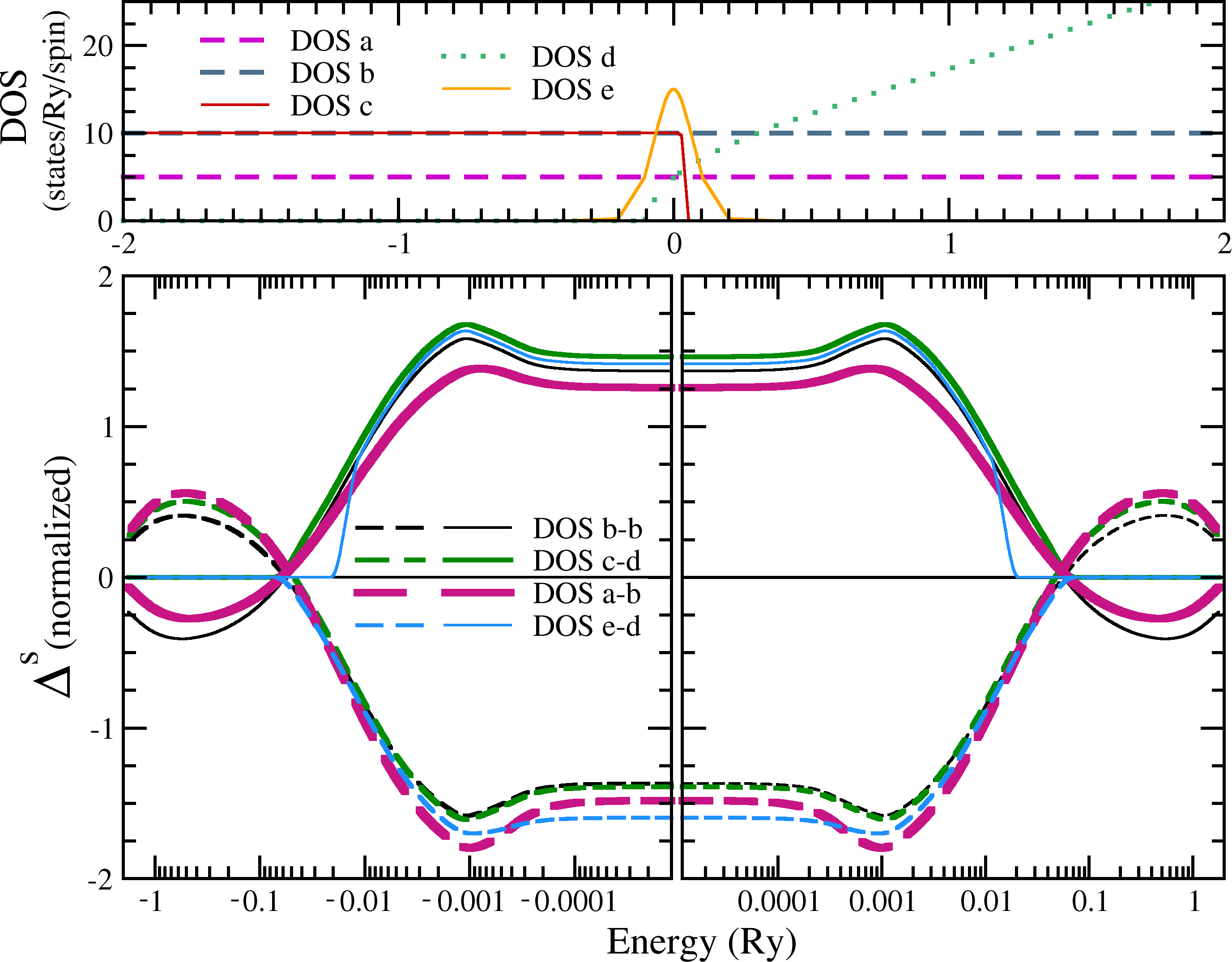} 
\par\end{centering}

\caption{Gap $\varDelta_{\pm}$ for different asymmetric density of states.
\label{fig:GapDos}}
\end{figure}

To verify this observation, we test different densities of states
instead of the constant one used so far: The different functions are
step and square root functions which represent a two and three dimensional
system, respectively and a Gaussian peak. The different functions
are shown in Fig.\ \ref{fig:GapDos}. The non flat functions cut
away the long energy tails of the gap function. However, the effect
on the critical temperature is rather small.

What has a strong effect on $T_{\text{c}}$ is a change of the ratio
$\frac{N_{+}\left(\epsilon_{\text{F}}\right)}{N_{-}\left(\epsilon_{\text{F}}\right)}=1$
(magenta line in Fig.\ \ref{fig:GapOfKappa}). This verifies that
in the two band system with a sign changing gap, only a small energy
region around the Fermi level matters for the Coulomb repulsion. This
is very different from the one band case, where the Coulomb renormalization
at large energies is an essential effect.

Last we consider the inclusion of the purely attractive phonon contribution.
It's behavior is rather straightforward. If a single phonon peak is
included ( Eq.~\ref{eq:ModelLambda} ) providing the same coupling
between all bands, the critical temperature reduces by increasing
$\lambda^{{\scriptscriptstyle \text{Ph}}}$. Until a the phononic
coupling strength reaches the value of $\lambda^{{\scriptscriptstyle \text{SF}}}$.
The phonons dominate the gap equation and the symmetry of the gap
changes. The $s_{\pm}$ state favored by the repulsive interactions
is suppressed and an $s_{++}$ state is found. From this point the
$T_{\text{c}}$ starts to rise again with increasing $\lambda^{{\scriptscriptstyle \text{Ph}}}$.

\section{Summary and Outlook}

In this work we have derived a fully \textit{ab-initio} effective
electron-electron interaction containing the effect of a pairing mediated
by spin-fluctuation. The derivation starts from many-body perturbation
theory and the introduction of a self-energy function, containing
the relevant diagrams originating from its vertex part, therefore
going beyond the $GW$ approximation. The vertex correction enter
the expression in the form of the particle-hole propagator, which
is a highly non-local object determined by a BSE. The solution of
the BSE would be computationally not feasible for realistic systems
instead, in Sec.\ \ref{sub:Local-Approximation}, we propose a local
approximation for the particle-hole propagator. In this limit the
equation for the self-energy becomes very transparent: spin-fluctuations
enter via the magnetic response functions, that can be calculated
effectively\cite{PhysRevLett.102.247206,PhysRevB.86.060412} within
linear response TD-DFT, and the coupling to the electrons is mediated
by the exchange-correlation kernel.

This effective interaction is in principle applicable to any theory
of SC, however in this work we cast it into the framework of SCDFT
by the construction of an explicit xc kernel (Sec.\ \ref{sec:Functional}).
In this way the full gap equation remains completely parameter free.

We show a first application of the new functional (Sec.~\ref{sec:TwoBandModel})
to a two band electron gas model. Application to real materials will
follow, however this further step needs the calculation of the magnetic
response function for the real system and will be the subject to further
investigation.

\bibliographystyle{apsrev4-1}
\bibliography{references}

\end{document}